\documentclass[12pt,a4paper]{article}
\pdfoutput=1
\usepackage{jheppub}
\usepackage{bm}
\usepackage{ifsym}
\usepackage{amsfonts}
\usepackage{bbm}
\usepackage{ccaption}
\usepackage{mathrsfs}
\usepackage{booktabs}
\usepackage{color}
\usepackage{wrapfig}
\usepackage{multicol}
\usepackage{multirow}
\usepackage{graphics}
\usepackage{doi} 
\usepackage{natbib,hyperref}

\usepackage{amsmath, amssymb, setspace}

\newcommand{\mathsym}[1]{{}}
\newcommand{\unicode}[1]{{}}

\newcommand{\beq}{\begin{equation}}		\newcommand{\enq}{\end{equation}}

\newcommand{\la}{\langle}
\newcommand{\ra}{\rangle}

\newcommand{\bx}{\mathbf{x}}
\newcommand{\by}{\mathbf{y}}
\newcommand{\bd}{\mathbf{d}}

\newcommand{\D}{\delta}
\newcommand{\w}{\omega}
\newcommand{\Dk}{ D k}
\newcommand{\Dp}{ D p}
\newcommand{\Dq}{ D q}

\newcommand{\vk}{\mathbf{k}}
\newcommand{\vp}{\mathbf{p}}
\newcommand{\vq}{\mathbf{q}}
\newcommand{\vr}{\mathbf{r}}
\newcommand{\vz}{\mathbf{s}}

\def\Df{\mathrm{D}}

  \captionnamefont{\bfseries}
  \captiontitlefont{\small\sffamily}
  \captiondelim{: }
  \hangcaption

\def\clock{{\count0=\time
           \divide\count0 60
           \ifnum\count0<10 0\fi\the\count0
           \multiply\count0 -60 \advance\count0 \time
           :\ifnum\count0<10 0\fi \the\count0
         }}
\newcommand{\timestamp}{{\small\vbox{\hbox{\tt\jobname.tex}
\hbox{\the\day/\the\month/\the\year, \clock}}}}

\usepackage{tocloft}

\definecolor{rust}{rgb}{0.8,0.2,0.2}
\definecolor{green}{rgb}{0.1,0.8,0.2}

\numberwithin{equation}{section}
\numberwithin{figure}{section}

\begin{document}

\begin{titlepage}
\title{A post-Gaussian approach to dipole symmetries and interacting fractons}
\author{J. Molina-Vilaplana}

\affiliation[]{Departamento de Autom\'atica, Universidad Polit\'ecnica de Cartagena,
C/Dr Fleming S/N, 30202, Cartagena, Spain.}

\emailAdd{javi.molina@upct.es}

\abstract{We use a post-Gaussian variational approach to non-perturbatively study a general class of interacting bosonic quantum field theories with generalized dipole symmetries and fractonic behaviour. We find that while a Gaussian approach allows to carry out a consistent renormalization group (RG) flow analysis of these theories, this only grasps the interaction terms associated to the longitudinal motion of dipoles, which is consistent with previous analysis using large $N$ techniques. Remarkably, our post-Gaussian proposal, by providing a variational improved effective potential, is able to capture the transverse part of the interaction between dipoles in such a way that a non trivial RG flow for this term is obtained and analyzed.  Our results  suggest that dipole symmetries that manifest due to the strong coupling of dipoles, may robustly emerge at low energies from short distance models without that symmetry.}


\maketitle
\flushbottom
\end{titlepage}

\newpage
\section{Introduction}
\label{sec:intro}
Fractons are exotic quasiparticles arising in lattice models which have raised a remarkable interest in both high energy physics and condensed matter~\cite{Chamon:2004,Haah:2011, Vijay:2015,Nand:2018,Pretko:2020,Slagle:2017, Seiberg:2020,Seiberg:2020b}. Continuous models exhibiting fractonic behaviour are realized through a specific class of field theories possessing a generalized notion of global symmetry known as subsystem symmetries and/or multipole symmetries. These are sensitive to the details of the underlying lattice and not compatible with Lorentz invariance \cite{McGreevy:2022}.  

Subsystem symmetry defines symmetry operators acting independently on subspaces of the total  system. The most simple examples of these models involve a single free real scalar field. As a fact, any interaction term allowed by this symmetry is irrelevant under renormalization, making these theories essentially free at low energies \cite{Seiberg:2020, Seiberg:2020b}.

A closed related symmetry is multipole symmetry, and concretely dipole symmetry, the type of symmetry in which we will focus in this work. In models with dipole symmetry, a U(1) charge and the associated dipole moment are conserved \cite{Pretko:2018}. Contrarily to models with subsystem symmetries, models with dipole symmetries can be made rotationally invariant.  Interacting models consistent with these symmetries are built using complex scalar fields. The resulting bosonic theories are characteristically non-Gaussian and strongly correlated as, in contrast to standard field theories, the only symmetry allowed terms with spatial derivatives in the Lagrangian have four or more powers of the fundamental fields \cite{Pretko:2018}. This makes these theories very hard to analyze through mean field or perturbative techniques. 

References \cite{Jensen:2022, Islam:2023} have studied large $N$ versions of models with conserved dipole moment in static and out equilibrium settings respectively. The large $N$ approach allows to study the physics of these systems at a finite interaction strength. While they obtained the Green's function (charge-charge propagator) of the models, the large $N$ factorization of higher correlation functions does not allow to compute the dipole propagator from which it is possible to fully access the non-Gaussian structure of the dipole-dipole interaction.

Regarding  renormalization group (RG) analysis of these models, it was pointed out in \cite{Li:2021} that  microscopic models with dipole symmetries may look quite unrealistic, highly anisotropic, and fine-tuned. There, authors found that adding as a perturbation, dipole-symmetry-disallowed kinetic terms to the type of large $N$ models commented above, the one-loop $\beta$-function shows that the UV model, which is not dipole-symmetric but a more realistic and less fine-tuned model, flows to the dipole-symmetric point  at the deep infrared limit, regarding the dipole-symmetry as an emergent phenomenon at low energies. In \cite{Distler:2021} a model of weakly interacting scalars with subsystem symmetries and non-relativistic fermions in 2+1 dimensions was analized in perturbation theory. It was shown that the 1-loop $\beta$-function of the fracton-fermion coupling does not flow due to an emergent symmetry of the effective Lagrangian. 
In \cite{Han:2022}, authors, using renormalization group analysis, investigated the emergence of  exotic non-Fermi liquids in two dimensions, in a model where fermions interact with a bosonic fracton system. In \cite{Grosvenor:2022}, a RG procedure was applied to a lattice model with fractonic behaviour and authors found that all screening effects generating non-transverse fractonic couplings vanish, thus indicating that dipole interactions might only exist transverse to the dipole direction. Indeed, they also found that  the coefficient of the transverse fractonic term did not get renormalized.

Given these results, it would be desirable to invoke a nonperturbative analysis of  models with dipole symmetries which, in addition to their manifest theoretical interest, are expected to be realized both experimentally and in nature \cite{Pretko:2017, Doshi:2020, Zechmann:2022}.

In this work, we carry out this analysis through a variational approach based on the functional Schr\"odinger picture in QFT \cite{Hatfield:1992, Symanzik:1981}. Being the Schrödinger representation not manifestly Lorentz invariant poses no particular problem here because we deal with theories whose symmetries are incompatible with Lorentz invariance. We use a method that allows to extend predictions for these systems beyond mean field theory and the Gaussian variational approch in a systematic way. Our method is based on nonlinear canonical transformations (NLCT) \cite{polley89, ritschel90, ibanez} from which it is possible to build  extensive variational wavefunctionals which are certainly non-Gaussian. 

Our results suggest that, in spite of all of the particularities of the models that we have addressed, our approach, provides useful insights into the physics of field theories with fractonic behaviour. Concretely, we find that a Gaussian approximation, while allowing to carry out a consistent RG flow analysis of the theory at 1-loop based on the Gaussian effective potential (GEP), only grasps the interaction terms associated to the longitudinal motion of dipoles. This agrees with the large $N$ approach taken in \cite{Jensen:2022, Islam:2023} and is not surprising, as the Gaussian variational approach coincides with the large $N$ solution for an interacting model. 

Remarkably, our proposal of non-Gaussian ansatz wavefunctional based on NLCT, by providing (-a variational approximation of-) the connected part of the four point function between charges (a.k.a dipole two point funcion) by means of an improved post Gaussian effective potential, is able to capture the transverse part of the the interaction between dipoles in such a way that a non trivial RG flow for this term is obtained and analyzed. This is something that cannot be derived from a pure Gaussian ansatz or large $N$ techniques. Namely, our non-perturbative calculation of the RG flow through its $\beta$-functions, suggests that dipole symmetries that manifest due to the strong coupling of dipoles, arise at low energies from less symmetric (in the dipole sense) UV models. 

The paper is organized as follows: in Section \ref{sec:model} we review some background on fractonic field theories with a dipole symmetry. In Section \ref{sec:gva} we carry out an analysis of these type of models through a variational Gaussian wavefunctional representing the ground state of the theory. Once a self-consistent solution of the variational parameters is provided, we carry out an analysis of the RG flow of the theory based on the Gaussian effective potential. Section \ref{sec:ng_approach} presents a post-Gaussian variational wavefunctional builded through non-linear canonical transformations (NLCT). Through the NLCT technique, we compute the connected part of the dipole-dipole interaction that contributes to improve the energy expectation value and the effective potential, from which we  study the RG flow of the model through the $\beta$-functions for the dipole-dipole interaction terms. We provide a final account of the work in Section \ref{sec:discussion}.

\section{Interacting Fracton Model}
\label{sec:model}
In this Section we review some background on field theories with dipole symmetry. A complex scalar field $\phi$ has been used to describe these theories, invariant both under global phase rotations $\phi\rightarrow e^{i\Lambda}\phi$ corresponding to conservation of charge, and also invariant under linear phase rotations, $\phi\rightarrow e^{i\, \bd\cdot \bx}\phi$ for constant $\bd$, corresponding to conservation of dipole moment \cite{Pretko:2018}. This is a special case of theories that poses a generalized notion of global symmetry  known as multipole symmetries (in this case, dipole) \cite{McGreevy:2022}.

\subsection{Dipole symmetry}
We consider a translational and rotationally invariant continuum quantum field theory in $(D+1)$ dimensions of a charged scalar field $\phi(t,\bx)$ which, in addition, it is also invariant under $U(1)$ phase rotations, and dipole transformations. That is, under a $U(1)$ phase rotation $\Lambda$ and dipole transformation $\bd$, the scalar field transforms as
\beq
    \phi(t,\bx) \to e^{i \Lambda - i\, \bd\cdot \bx}\, \phi(t,\bx)\,.
\enq
In order to build an action invariant under this transformation it is useful to find covariant operators, that is, operators that are invariant under this transformation up to a phase factor \cite{Pretko:2018}.  The result is that, in contrast to standard field theories, these theories do not possess any covariant operators featuring spatial derivatives, that is, acting on only a single $\phi$ operator.  Instead, as it has been shown in \cite{Pretko:2018},  the lowest order covariant spatial derivative operator contains two factors of $\phi$, taking the form:
\beq\label{eq:bilocal_D}
    \Df_{ij}(\phi,\phi) = \left( \phi\partial_i \partial_j \phi - \partial_i \phi \partial_j \phi \right)\, ,
\enq
which transforms covariantly as $\Df_{ij}(\phi,\phi) \to e^{2 (i\Lambda-\bd\cdot \bx)} \Df_{ij}(\phi,\phi)$; see~\cite{Pretko:2018}. Using these covariant operators, we can then write down a lowest order action for this theory as:
\beq\label{eq:model_pretko}
    S = \int dt\, d^{D} \bx\,  \left( 
     |\partial_t \phi|^2 
    - \lambda\,  \Df_{ij}(\bar{\phi}, \bar{\phi}) 
    \Df^{ij}(\phi,\phi)
    - U(\bar{\phi}\, \phi)\right)\, ,
\enq
where $\lambda$ is a dipole-dipole coupling constant and $U(\bar{\phi}\, \phi)$ is an interaction potential term for the isolated charges.  We note that the usual spatial kinetic term $\partial_i\bar{\phi}\, \partial^i\phi$ is forbidden by dipole symmetries, and the simplest terms with spatial derivatives involve at least four powers of the charged field written above. 

For the sake of convenience in the rest of this paper, we note that dipole symmetry is simpler to analyze in momentum space. Namely, a dipole transformation acts on $\phi(t,\vk)$ as
\beq
    \phi(t,\vk) \to \phi(t,\vk- \bd)\,,
\enq
and on the conjugate field by the opposite shift, that is, as translations in momentum space. Thus, dipole-invariant interactions must be invariant under translations in momentum space and therefore they must only depend on momentum differences. As an illustration, one may write the interaction $\Df_{ij}(\phi^{\dagger},\phi^{\dagger})\Df^{ij}(\phi,\phi)$  in momentum space as \cite{Jain:2023}
\beq\begin{split}\label{eq_premodel}
    \lambda\, \int dt\,  d^{D} \bx\, 
    & \Df_{ij}(\bar{\phi},\bar{\phi})\, 
    \Df^{ij}(\phi,\phi)  \\ 
    &= \int dt\,  \Dk_1\, \ldots \Dk_4 \, 
    V^{(4)}(\vk_1,\vk_2,\vk_3,\vk_4)\,
    \bar{\phi}(t,\vk_1)\,
    \bar{\phi}(t,\vk_2)\,
    \phi(t,\vk_3)\, 
    \phi(t,\vk_4)\,,
\end{split}
\enq
where we use the notation $\int Dk \equiv \int d^{D}\vk/(2\pi)^{D}$ and the vertex is defined as
\beq
    V^{(4)}(\vk_1,\vk_2,\vk_3,\vk_4) = \frac{\lambda}{4}
    \left[ \vk_{12}\cdot \vk_{34}\right]^2 
    \delta^{D}(\vk_1 +\vk_2+ \vk_3 + \vk_4),
\enq
 with $\vk_{mn} = \vk_m - \vk_n$.
\subsection{Model of interacting Fractons with dipole-symmetry}
In this work, following \cite{Pretko:2018} and \cite{Jensen:2022} we consider a more "sophisticated" model than \eqref{eq_premodel} given by the action
\beq\label{eq:fracton_lagrangian}
	S = \int dt\, d^{D} \bx\,  \Bigg( |\partial_t\phi|^2 - m^2|\phi|^2 -\lambda_T \Df_{\{ij\}}(\phi,\phi)\Df^{\{ij\}}(\bar{\phi},\bar{\phi}) - \lambda_S\, \left| \delta^{ij}\Df_{ij}(\phi,\phi)+ \gamma |\phi|^2\right|^2\Bigg)\,,
\enq
where 
\beq
\Df_{\{ij\}}(\phi,\phi) = \Df_{ij}(\phi,\phi) - \frac{\delta_{ij}}{D}\delta^{kl}\Df_{kl}(\phi,\phi)\, ,
\enq
 is the traceless part of $\Df_{ij}(\phi,\phi)$ and $\gamma$ is a complex parameter. A mass term is included which have no interplay with the spatially dependent phase rotation. The constants $\lambda_T$ ($T$ is for tensor) and $\lambda_S$ ($S$ is for scalar) are arbitrary couplings describing the longitudinal motion of dipoles($S$) and the transverse one ($T$) respectively. This model posses a characteristic non-Gaussian form and thus it is hard to deal with it.

In this work we use variational approaches through the functional Schr\"odinger picture in QFT \cite{Hatfield:1992, Symanzik:1981}. The functional Schr\"odinger picture works with the Hamiltonian of the theory in \eqref{eq:fracton_lagrangian}, which, written in momentum space reads as \cite{Jensen:2022}
\beq\begin{split}\label{eq:hamilt}
H &= \frac{1}{2}\, \int \Dk \left[\bar{\pi}(\vk) \pi(-\vk) + m^2 \bar{\phi}(\vk) \phi(-\vk)\right]\\
&  + \int \Dk_1 \Dk_2 \Dk_3 \Dk_4 \, V^{(4)}(\vk_1,\vk_2,\vk_3,\vk_4)\,  \bar{\phi}(\vk_1)\bar{\phi}(\vk_2)\phi(\vk_3)\phi(\vk_4)\, \delta^{D}(\vk_1 +\vk_2+ \vk_3 + \vk_4)\, ,
\end{split}\enq
where the conjugate momentum of the field $\phi(\vp)$ is $\pi(\vp)\equiv-i\delta/\delta\phi(-\vp)$, in such a way that $[\phi(\vp), \pi(\vq)] = i\delta^{D}(\vp + \vq)\,$ and the interaction vertex between dipoles is given by
\beq\label{eq:vertex}
V^{(4)}(\vk_1,\vk_2,\vk_3,\vk_4) = \frac{\lambda_T}{4}\left[ \left(\vk_{12}\cdot \vk_{34}\right)^2 - |\vk_{12}|^2 |\vk_{23}|^2\right] +  \frac{\lambda_S}{4}\left[ \left(|\vk_{12}|^2 + 2\bar{\gamma}\right)\left(|\vk_{34}|^2 + 2\gamma\right)\right]\, .
\enq

The Hamiltonian \eqref{eq:hamilt} is bounded from below for $\lambda_T\geq 0$ and $\lambda_T + D\,  \lambda_S \geq 0$ \cite{Bidussi:2021}. In this theory, the only allowed processes  are those in which the total dipole moment is conserved. Hence, isolated charges have restricted mobility and thus show fractonic behaviour. Remarkably, the non-Gaussian dipole-dipole interactions do not forbid a fractonic charge to move completely as far as the (restricted) charge mobility arises from  processes in which the total dipole moment is conserved. Namely, the locality of dipole-dipole interactions in \eqref{eq:hamilt} forces that such processes must be mediated by a propagating dipole between the two charges \cite{Pretko:2020,Pretko:2018}.  

Given this qualitative expected behaviour, in this work we will mainly be focused on providing some quatitative responses on \emph{i)} the characterization of the dipole dispersion relation and \emph{ii)} using standard renormalization group (RG) analysis, investigate how and to what extent it can be considered that the dipole-symmetry is an emergent phenomenon at low energies. In the next Sections, we investigate on this topics by invoking variational nonperturbative techniques.

\section{Gaussian Variational Approach}
\label{sec:gva}
In this Section we carry out an analysis of the theory given by the Hamiltonian in \eqref{eq:hamilt} through a variational approach using the functional Schr\"odinger picture in QFT \cite{Hatfield:1992, Symanzik:1981}. Being the Schrödinger representation  not manifestly Lorentz invariant, poses no particular problem here as we tackle with theories whose multipole and subsystem symmetries are incompatible with Lorentz invariance. Therefore we use a Gaussian variational wavefunctional representing the ground state of the theory. By the variational method we provide a self-consistent solution of the variational parameters and carry out a RG flow analysis of the theory based on the Gaussian Effective Potential \cite{Coleman:1973, Barnes:1980}.

\subsection{Gaussian wavefunctional}
\label{sec:wf_gva}
The variational Gaussian wavefunctional that we take as an ansatz for the ground state of the theory \eqref{eq:hamilt} is given by
\beq\label{eq:gauss_ansatz}
\Psi_G\left[\phi, \bar{\phi}\right]= N\, \exp\left[-\frac{1}{2}\int \Dk \Dk'\, \bar{\phi}(\vk)\, G^{-1}(\vk)\, \phi(\vk') \delta^{D}(\vk + \vk')\right]\, ,
\enq
where the variational kernel $G(\vk)$ satisfies
\beq\begin{split}
\langle \bar{\phi}(\vk)\, \phi(\vk') \rangle &= G(\vk)\, \delta^{D}(\vk + \vk')\\
\langle \bar{\pi}(\vk)\, \pi(\vk') \rangle &= \frac{1}{4} G^{-1}(\vk)\, \delta^{D}(\vk + \vk')\, ,
\end{split}
\enq
and $N = [{\rm det}(2\pi G)]^{-1/2}$. As it will be commented below, the real space representation of $G(\vk)$,  $G(\bx-\by) =\langle \bar{\phi}(\bx)\, \phi(\by) \rangle$, may be interpreted as a the one point function of a dipole creation operator, that is, an operator creating a charge at $\by$ and anti-charge at $\bx$ and thus it can be used as a dipole order parameter \cite{Jensen:2022}.

\subsection{Ground state energy density}
\label{sec:energy_gva}
The energy density of the ground state, using Wick's theorem, is given by
\beq\begin{split}\label{eq:energy}
\mathcal{E}&=\frac{1}{\rm Vol} \langle \Psi_G\vert H\vert \Psi_G\rangle= \int \Dk\,  \frac{1}{8} G^{-1}(\vk) + \frac{1}{2}\, m^2\, G(\vk) 
+ \int \prod\, \Dk_i \, \delta^{D}(\sum \vk_i)\, V^{(4)}(\vk_1,\vk_2,\vk_3,\vk_4)\,  \\
& \times \Bigg[ G(\vk_1) G(\vk_2) \delta^{D}(\vk_1+\vk_3)\delta^{D}(\vk_2+\vk_4)  
+ G(\vk_1) G(\vk_2) \delta^{D}(\vk_1+\vk_4)\delta^{D}(\vk_2+\vk_3)\Bigg] \\
& = \int \Dk\, \frac{1}{8} G^{-1}(\vk) + \frac{1}{2}\, m^2\, G(\vk)
+ 2\, \int \Dk \Dk'\, \bar{V}^{(4)}(\vk,\vk')\,  G(\vk) G(\vk')\, ,
\end{split}
\enq
where ${\rm Vol}\equiv \delta^{D}(0)$ and in the last line we have used that
 $\bar{V}^{(4)}(\vk,\vk')\equiv V^{(4)}(-\vk,-\vk',\vk,\vk')=V^{(4)}(-\vk,-\vk',\vk',\vk)$ with 
\beq\label{eq:fordw_scatt}
\bar{V}^{(4)}(\vk,\vk') = \frac{\lambda_S}{4}\vert |\vk -\vk'|^2 + 2\gamma\vert^2\, ,
\enq
 being the forward scattering limit of $V^{(4)}(\vk_1,\vk_2,\vk_3,\vk_4)$ \cite{Jensen:2022}. This is in accordance with  the Gaussian ansatz being a consistent truncation of the Dyson series. In fact, this imposes that the model is consistent only when the forward limit is positive, i.e, $\lambda_S >0$. With this, one may write
\beq\begin{split}\label{eq:energy_variational}
\mathcal{E}[G]= \int \Dk\, \Bigg[\frac{1}{8} G^{-1}(\vk) + \frac{1}{2}\, m^2\, G(\vk) + \Sigma(\vk) G(\vk)\Bigg]\, ,
\end{split}
\enq
after defining the self-energy $\Sigma(\vk)$ as
\beq\label{eq:self_energy}
\Sigma(\vk)=2\int \Dk'\,\bar{V}^{(4)}(\vk,\vk')\, G(\vk')\, . 
\enq

The variational parameter $G(\vk)$ is thus obtained by
\beq\label{eq:variation}
\frac{\D \mathcal{E}[G]}{\D G(\vk)} = 0
\enq
which yields the truncated Dyson equation (\emph{gap equation})
\beq\label{eq:Dyson}
-\frac{1}{4}G^{-2}(\vk) + m^2 + 2\, \Sigma(\vk)=0\, ,
\enq
and thus
\beq
G(\vk) = \frac{1}{2\,\w(\vk)}\, ,\quad \w^2(\vk) = 2\, \Sigma(\vk) + m^2\, .
\enq

Several comments are due here. 
\begin{enumerate}
\item The variational principle and the Gaussian form of the variational wave functional \eqref{eq:gauss_ansatz} implies a gap equation that determines the Green function $G(\vk)$ in the static case. 

\item The variational Green function $G(\vk)$ acts as order parameter for dipole breaking: if the position space Green's function $\langle \bar{\phi}(\bx)\, \phi(0)\rangle\nsim \delta^{D}(\bx)$, then the dipole symmetry acts on it. In other words: $G$ can be understood as the one-point function of an operator that creates a dipole, with a charge at $0$ and an anticharge at $\bx$. Thus, being $G(\vk)$ nonzero, it is understood as a dipole condensate in analogy with the usual charge condensate associated with ordinary symmetry breaking \cite{Jensen:2022}.

\item We note that the dipole symmetry acts on the momentum space field by $\phi(\vk) \to\phi(\vk - \bd)$, i.e. by a shift of momentum, and on the conjugate field by the opposite shift. Dipole transformations then act on $G(\vk)$  by $G(\vk)\to G(\vk + \bd)$ and similarly for $\Sigma(\vk)$ as $\bar{V}^{4)}(\vk,\vk')$ only depends on the difference of momenta, that is, under a dipole transformation
\beq\begin{split}
2\int \Dk'\, \bar{V}^{4)}(\vk,\vk') G(\vk'+\bd) &= 2\int \Dq\, \bar{V}^{4)}(\vk,\vq-\bd) G(\vq)\\
&= 2\int \Dq\, \bar{V}^{4)}(\vk +\bd,\vq) G(\vq) = \Sigma(\vk + \bd)\, .
\end{split}
\enq

\item For infinite volume, the theory in Eq \eqref{eq:fracton_lagrangian} posses a continuous non-compact dipole symmetry $\mathbb{R}^D$. This implies the existence of a family of solutions  $\Psi_G^{\bd}\left[\phi, \bar{\phi}\right]$ with the same energy, labeled by allowed dipole transformations. In a lattice regularization, $\bd$ will be valued in the space of momenta, that is the Brillouin zone. Thus, the number of allowed dipole transformations, and so $\Psi_G^{\bd}\left[\phi, \bar{\phi}\right]$ solutions related by dipole symmetry, would be then equal to the number of lattice sites $N_{\rm sites} \sim {\rm Vol}\, \cdot V_{\rm BZ}$, with $V_{\rm BZ}$ the (-dimensionless-) volume of the Brillouin zone \cite{Jensen:2022}.

\item  As commented above, the dipole symmetry is lattice sensitive, and thus the theory \eqref{eq:fracton_lagrangian} exhibits features of UV/IR mixing characteristic of fracton models. Nevertheless, as stated in \cite{Jain:2023}, the UV-sensitivity showed by these models, is mild in comparison with models exhibiting subsystem symmetries. As discussed in \cite{Seiberg:2021}, for the later,  the UV/IR mixing refers to the low-energy mixing among small and high momenta. In other words, depending on the chosen direction, the low-energy modes can have very high momenta (see Appendix \ref{sec:appA}). As a consequence it has been suggested that renormalization group is not applicable to models exhibiting fractonic behaviour \cite{Zhou:2021}. However, it has been argued that this can be addressed by conforming the integration of the high-energy modes to the symmetries of the
fracton models \cite{Lake:2021}.

\item In a theory without interactions, $G(\vk) \propto 1/m$, that is, there are no propagating charges which amounts to a momentum dependent self-energy $\Sigma(\vk)=0$. Instead, interactions imply $\Sigma(\vk)\neq 0$ and thus allow restricted charge mobility. The precise nature of this mobility can be established by finding self consistent solutions for $\Sigma(\vk)$. Unlike the case of subsystem symmetries, $\Sigma(\vk)$ in \eqref{eq:self_energy} is rotationally invariant. Indeed, following \cite{Jensen:2022}, we note that the vertex $\bar{V}^{(4)}(\vk,\vq)\sim \vert |\vk -\vq|^2 + 2\gamma \vert^2$ is a polynomial in $|\vk|$ of degree 4, and thus, according to \eqref{eq:self_energy}, so $\Sigma(\vk)$ is too. Therefore, we make a rotationally invariant ansatz 
\beq
\Sigma(\vk) = a_0 + a_1 |\vk|^2+a_2 |\vk|^4\, ,
\enq
with parameters $(a_0, a_1, a_2)$ to be self-consistently determined. In Appendix \ref{sec:appB}, we find these solutions numerically.

\item From Eqs \eqref{eq:fordw_scatt} and \eqref{eq:energy_variational}, one realizes that the Gaussian approximation to the ground state only "sees" the $S$-term of the dipole-dipole interaction, as only captures the forward scattering limit of \eqref{eq:vertex}. This result agrees with the large $N$ approach taken in \cite{Jensen:2022, Islam:2023} and is not surprising, as it is well known that the Gaussian ansatz coincides with the large $N$ solution for an interacting model. An obvious question is if there is a systematic way to improve the ansatz going beyond the Gaussian regime and if this procedure is capable to grasp the transverse part of the interaction vertex. This will be addressed in the next Section but before, we will analyze further the Gaussian solution.
\end{enumerate}

\subsection{A more general Gaussian wavefunctional}
\label{sec:gauss_condensate}
For subsequent developments, it is worth to consider a more general Gaussian ansatz. To this end, we consider a wavefunctional that accounts for the existence of a nonzero charge condensate,
\beq\label{eq:SG_Gaussian}
\Psi_{SG}\left[\phi, \bar{\phi}\, \right]= N\, \exp\left[-\frac{1}{2}\int \Dk \Dk'\,  \left(\bar{\phi}(\vk) - \bar{\sigma}\right)\, G^{-1}(\vk) \left(\phi(\vk')-\sigma\right) \delta^{D}(\vk + \vk')\right]\, ,
\enq
and which can be obtained from $\Psi_{G}\left[\phi, \bar{\phi}\, \right]$ in \eqref{eq:gauss_ansatz} by
\beq\label{eq:sg_transform}
\Psi_{SG}\left[\phi, \bar{\phi}\, \right]
=
\exp(O_S)\Psi_{G}\left[\phi, \bar{\phi}\, \right]
\ ,
\enq
with
\beq
O_S
=
-\int \Dp\, \left(\sigma\, \frac{\delta}{\delta\phi(-\vp)} + \bar{\sigma}\, \frac{\delta}{\delta\bar{\phi}(\vp)}\, \right)\, \delta^{D}(\vp)
\ ,
\enq

Operationally, expectation values w.r.t $\Psi_{SG}$ amount to expectations values w.r.t $\Psi_{G}$ under the substitutions
\beq\begin{split}
\la \bar{\phi}(\vk) \phi(-\vk)\ra &\to G(\vk)\,  +  \, \sigma^{2}\, \delta^{D}(\vk)\\
\la \bar{\pi}(\vk) \pi(-\vk)\ra &\to \frac{1}{4} G^{-1}(\vk)\, ,
\end{split}
\enq

Under this ansatz, the energy density reads
\beq\label{eq:energy_variational_condensate}
\mathcal{E}=\int \Dk\, \Bigg[\frac{1}{8} G^{-1}(\vk) + \frac{1}{2}\, m^2\, \left(G(\vk)\,  +  \, \sigma^{2}\, \delta^{D}(\vk)\right) + \Sigma_{\sigma}(\vk) \left(G(\vk)\,  +  \, \sigma^{2}\, \delta^{D}(\vk)\right)\Bigg]\, ,
\enq
with $\Sigma_{\sigma}(\vk)$ given by
\beq\label{eq:self_energy_sigma}
\Sigma_{\sigma}(\vk)=
2\int \Dq\, \bar{V}^{(4)}(\vk,\vq)\, \left(G(\vq)\,  +  \, \sigma^{2}\, \delta^{D}(\vq)\right)
= \Sigma(\vk)  + 2\, \sigma^{2}\, \bar{V}^{(4)}(\vk,0)\,\, .
\enq

We note that in \eqref{eq:energy_variational_condensate}
\beq\begin{split}\label{eq:energy_variational_condensate_term_a}
\int \Dk\, \Sigma_{\sigma}(\vk) G(\vk)\,  +  \int \Dk\, \sigma^{2}\, \Sigma_{\sigma}(\vk)\, \delta^{D}(\vk) 
= \int \Dk\, \Sigma_{\sigma}(\vk) G(\vk)  + \sigma^{2}\, \Sigma_{\sigma}(\vk = 0)\\
= \int \Dk\, \Sigma_{\sigma}(\vk) G(\vk)  + \sigma^{2}\, \left[\Sigma(\vk = 0) + 2\sigma^{2}\bar{V}^{(4)}(0,0)\right] = \int \Dk\, \Sigma_{\sigma}(\vk) G(\vk)  + 2\sigma^{4}\, \gamma^{2}\, ,
\end{split}
\enq
where in the last equality, we used that  $\bar{V}^{(4)}(0,0)=  \lambda_S\, \gamma^2$ and we imposed
\beq
\sigma\, \Sigma(\vk=0) = 0\, .
\enq
For $\sigma \neq 0$, this single condition imposes that $\Sigma(\vk)$ to be gapless at $\vk=0$. Contrarily, if $\Sigma(\vk)$ is gapped, the system has to be such that $\sigma=0$ ($U(1)$ is unbroken). In other words, this single restriction allows to consider different phases of the model as it is explained in Appendix B, \cite{Jensen:2022}. 

With this, the energy density now reads as 
\beq\label{eq:energy_condensate}
\mathcal{E}=\int \Dk\, \Bigg[\frac{1}{8} G^{-1}(\vk) + \frac{1}{2}\, m^2\, G(\vk) + \Sigma_{\sigma}(\vk) G(\vk)\Bigg] + \frac{1}{2} m^{2}\, \sigma^{2} + 2\sigma^{4}\,  \lambda_S\, \gamma^{2}\, ,
\enq

and the variational procedure yields the truncated Dyson equation (gap equation),
\beq\label{eq:Dyson_sigma}
-\frac{1}{4}G^{-2}(\vk) + m^2 + 2\Sigma_{\sigma}(\vk)=0\, ,
\enq
from which
\beq\label{eq:Sigma_condensate}
G(\vk) = \frac{1}{2\,\w(\vk)}\, ,\quad \w^2(\vk) = 2\, \Sigma_{\sigma}(\vk) + m^2\, .
\enq

\subsection{Renormalization and $\beta$-function}
\label{sec:renormalization_gva}
Following \cite{Barnes:1980}, we define the Gaussian Effective Potential (GEP) $\mathbb{V}_{\rm eff}[\sigma]\equiv \mathcal{E}$ from \eqref{eq:energy_condensate} as 
\beq\begin{split}\label{eq:gep}
\mathbb{V}_{\rm eff}[\sigma] &= \frac{1}{4}\, \int \Dk\, \left[2\, \Sigma_{\sigma}(\vk) + m^{2} \right]^{1/2}\, +\frac{1}{2} m^{2}\, \sigma^{2} + 2\, \lambda_S\, \gamma^2\, \sigma^{4}\, ,
\end{split}
\enq

The GEP usually contains divergences only because it is written in terms of bare parameters
$m$ and  $\lambda_S$. Once reexpressed in terms of renormalized parameters $m_R$ and $\lambda_R$,  $\mathbb{V}_{\rm eff}[\sigma]$ becomes manifestly finite. This reparametrization of the theory, or "renormalization", can not change the physical content of the theory. That is, choosing different renormalized parameters $m_R^{'}$ and $\lambda_S^{'}$ would
lead to a different looking but equivalent $\mathbb{V}_{\rm eff}[\sigma]$. As noted in \cite{Stevenson:1985}, the GEP is exactly renormalization-group (RG) invariant, which means that all ways of defining the renormalized parameters are equivalent, and our task reduces to merely finding the most convenient one. Noteworthily, this is not the case for the effective potential computed at one-loop, for which the RG invariance is spoiled by a "renormalization-scheme-dependence problem", just as in perturbation theory. 

\paragraph*{Renormalized mass.}
From the Gaussian effective potential \eqref{eq:gep}, one may define the renormalized mass $m_{R}$ as
\beq\label{eq:renorm_mass}
m_{R}^2 = 2\, \frac{d\, \mathbb{V}_{\rm eff}}{d \sigma^{2}}\Bigg|_{\sigma=0}\, ,
\enq

which gives
\beq\label{eq:renorm_mass}
m_{R}^2 = m^{2} +  \int \Dk\, \bar{V}^{(4)}(\vk,0)\, \left(2\, \Sigma_0(\vk) + m^{2}\right)^{-1/2}= m^2 + 2\, I_1(0) \, \,
\enq
where we have introduced the convenient notation 
\beq
I_N(M) = \frac{1}{2}\, \int \Dk\, \left[\bar{V}^{(4)}(\vk,0)\right]^N\, \left(2\, \Sigma_M(\vk) + m^{2}\right)^{1/2-N}\, .
\enq
The renormalized parameter $m_R$, as defined above, is very convenient because it turns out to be the mass of a one-particle excitation in the $\sigma=0$ vacuum \cite{Barnes:1980}. It is a finite quantity once the momentum integral is dimensionally regularized as explained in Apendix \ref{sec:appB}. 

\paragraph*{Renormalized coupling.} The renormalized coupling constant is defined by
\beq
\lambda_R = \frac{1}{4!}\frac{d^{4}\, \mathbb{V}_{\rm eff}}{d \sigma^{4}}\Bigg|_{\sigma=0} = \frac{1}{2}\, \frac{d^{2}\, \mathbb{V}_{\rm eff}}{d (\sigma^{2})^2}\Bigg|_{\sigma=0}\, ,
\enq
and is given by,
\beq\label{eq:renorm_coupling}
\lambda_R = \lambda_B - I_2(0) = \lambda_B \left[1-\lambda_B\, \mathcal{I}_2(0)\right]
\enq
where $\lambda_B= 2 \gamma^2\, \lambda_S = 2\, \bar{V}^{(4)}(0,0)$ and
\beq
\mathcal{I}_2(M) = \frac{1}{128}\, \int_{q}\, \left(\frac{|\vq^2 + 2\gamma|}{\gamma}\right)^4\, \left(2\, \Sigma_M(\vq) + m^{2}\right)^{-3/2}\, .
\enq

With the renormalized parameters \eqref{eq:renorm_mass} and \eqref{eq:renorm_coupling}, it would be possible to write a finite expression for $\mathbb{V}_{\rm eff}[\sigma]$. This would be quite convenient in order to study the phase diagram of the theory and its critical points.  We leave for a future investigation the use of the GEP to study the phase diagram of the models under consideration  while focusing the scope of this work on knowing how the coupling constant changes with the energy scale. 

To this end, now following Coleman and Weinberg \cite{Coleman:1973} and Barnes and Ghandour \cite{Barnes:1980}, we define the renormalized coupling as
\beq
\lambda_M = \frac{1}{4!}\frac{d^{4}\, \mathbb{V}_{\rm eff}}{d \sigma^{4}}\Bigg|_{\sigma=M} = \frac{1}{2}\, \frac{d^{2}\, \mathbb{V}_{\rm eff}}{d (\sigma^{2})^2}\Bigg|_{\sigma=M}\, ,
\enq
where the mass scale $M$ is nonzero but arbitrary. The result yields
\beq\label{eq:renormalized_coupling}
\lambda_M = \lambda_B - I_2(M) = \lambda_B \left[1- \lambda_B\, \mathcal{I}_2(M)\right]\, .
\enq
As $M$ is merely an arbitrary substraction point \cite{Coleman:1973}, it is non-physical and thus, no physical quantity can depend on it. In particular, for $\mathbb{V}_{\rm eff}$, one may write the RG-flow equation
\beq
M\, \frac{d\, \mathbb{V}_{\rm eff}}{d\, M}=\left(M\, \frac{\partial}{\partial M} + M\, \frac{\partial \, \lambda_M}{\partial M}\, \frac{\partial}{\partial \lambda_M}\right)\, \mathbb{V}_{\rm eff}=0\, .
\enq

This implies that we may define a $\beta$-function as
\beq\label{eq:beta_func}
\beta = M\, \frac{\partial \, \lambda_M}{\partial M}\, ,
\enq
which reads as 
\beq\label{eq:beta_func_B}
\beta = 3 M^2\, I_3(M)= 3\, \lambda_B^3\, \Lambda(M)\, ,
\enq
with  $\Lambda(M) \equiv M^2\, \mathcal{I}_3(M)$, and 
\beq
\mathcal{I}_3(M) = \frac{1}{128}\, \int_{q}\, \left(\frac{|\vq^2 + 2\gamma|}{\gamma}\right)^6\, \left(2\, \Sigma_M(\vq) + m^{2}\right)^{-5/2}\, .
\enq

It is convenient to write $\beta=\beta(\lambda_M)$. To this end we replace $\lambda_B$ with $\lambda_M$ everywhere except for first order terms. This substitution is valid to 1-loop, since it induces changes only for higher loops \cite{Hatfield:1992}. That is, we may invert Eq. \eqref{eq:renormalized_coupling} to solve for $\lambda_B$ in terms of $\lambda_M$ to obtain 
\beq
\lambda_B=\lambda_M  + \lambda_M^2\, \mathcal{I}_2(M)\, .
\enq
Then substituting into \eqref{eq:beta_func_B} one obtains the  $\beta$-function at 1-loop
\beq\label{eq:beta_func_M}
\beta(\lambda_M) = 3\, \lambda_M^3\, \Lambda(M) + \mathcal{O}(\lambda_M^5)\, .
\enq 
To debunk the exact behaviour of $\lambda_M$ on the energy scale  $M$, requires to exactly determine $\Lambda(M)$. This can only be done numerically, using the methods described in Appendix \ref{sec:appB}. In Figure \eqref{fig:LambdaM} we show the behaviour of $\Lambda(M)$ for $D=3$. To this end, $\Sigma_{M}(\vk)$ in \eqref{eq:self_energy_sigma} is numerically determined for different values of $M$. The Figure \eqref{fig:LambdaM} shows a positive function of the energy scale $M$. For the case shown in the figure, there is no fixed point $\lambda_F$ for which $\beta(\lambda_F)=0$. This means that \eqref{eq:beta_func_M} is always positive and thus implies that the $S$-part of the interaction vertex grows towards the UV from an arbitrary bare inital value. 
\begin{figure*}
    \centering
    \includegraphics[width = 0.4\linewidth]{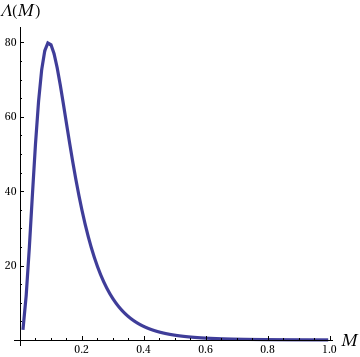}
    \caption{$\Lambda(M)$ in $D=3$, with $m = 0.1$ and $\gamma=0.1$.}
    \label{fig:LambdaM}
\end{figure*}

\section{Post-Gaussian Variational Approach}
\label{sec:ng_approach}
In this Section we present a post-Gaussian variational wavefunctional builded through non-linear canonical transformations (NLCT) \cite{polley89, ritschel90, ibanez}. The goal is the following: as shown above,  the Gaussian ansatz only captures the \emph{longitudinal/scalar} part of the interaction but obviates the \emph{transverse/tensor} part. Here, using a concrete class of post-Gaussian wavefunctionals generated through the NLCT technique, we compute the connected part of the dipole-dipole interaction (which amounts to a connected part of a four point function between charges). From this, it is possible to directly compute many interesting physical properties of the model \eqref{eq:fracton_lagrangian} which are hidden for both  the Gaussian ansatz and the large $N$ approach. 

Before going into this, it is worth to highlight some relevant issues related with applying a variational method in QFT \cite{Kogan:1994, Melgarejo:2020,Melgarejo:2020b}. To study the the most relevant features of the ground state of the theory, the choosen variational states must posses \emph{nice} calculability properties. That is to say, in order to first optimize and then compute physical properties of the theory, one needs to  evaluate expectation values of operators such as $n$-point functions. Here the problem is obvious as evaluations of expectation values of these quantities with respect to general non-Gaussian states are highly limited. In practice, this fact has  restricted the use of variational wavefunctionals to the Gaussian case. This poses a severe limitation to explore truly nonperturbative effects. For instance, Gaussian wavefunctionals, which amount to a set of decupled modes in momentum space, cannot describe the prototypical interaction between the high and low momentum modes of an interacting QFT, not to say the interplay between these modes in theories with the strong UV/IR mixing effects of those that we are considering in this work. For this reason, a feature that one should require for a variational wavefunctional is to include parameters describing the interplay between the high energy and low energy modes and how the former affects the low energy physics of our theory.

In the following, such a class of variational ansatz is used to study the theory in \eqref{eq:fracton_lagrangian}.

\subsection{Non-Gaussian states through NLCT}
\label{sec:nlct}
Following \cite{polley89, ritschel90, ibanez}, a class of extensive non-Gaussian variational wavefunctionals can be non-perturbatively built as\footnote{One could consider a sequence of perturbatively built trial states generated by polynomial corrections to a Gaussian state. However, since polynomial corrections are related to a finite number of particles, the associated effects are suppressed in the thermodynamic limit.}
\beq\label{eq:NG_wavefunc}
\Psi_{NG}[\phi,\,  \bar{\phi}] =  \mathcal{U}\, \Psi_G[\phi,\,  \bar{\phi}]= \exp(B)\, \Psi_G[\phi,\,  \bar{\phi}]\, ,
\enq
where $\Psi_G \equiv \Psi_G[\phi,\, \bar{\phi}]$ is the normalized Gaussian wavefunctional in Eq. \eqref{eq:gauss_ansatz} (- or equivalently \eqref{eq:SG_Gaussian}-) and 
\beq\label{eq:ng_transform}
\mathcal{U}\equiv\exp(B)\, ,
\enq
 being $B$, an anti-Hermitian operator ($B^{\dagger} = -B$) with variational parameters different to those characterizing the Gaussian wavefunctional $\Psi_G \equiv \Psi_G[\phi,\, \bar{\phi}]$. As it will be made explicit below, Eq. \eqref{eq:NG_wavefunc} can be understood as a non-linear version of the transformation in \eqref{eq:sg_transform}. 
 
The functional structure of \eqref{eq:ng_transform} implies that the expectation value of an operator $O$, that in general depends on $\phi,\, \bar{\phi},\,  \pi,\,  \bar{\pi}$, with respect to $\Psi_{NG}$  reduces to the computation of a Gaussian expectation value of the transformed operator $O \to  \mathcal{U}^{\dagger}\, O\, \mathcal{U}$. The relevant poit here is that concrete and suitable forms of $B$, while leading  to  non-Gaussian wavefunctionals  \eqref{eq:NG_wavefunc}, automatically truncate after the first non-trivial term, the infinite nested commutator series
\beq
\mathcal{U}^{\dagger}\, O\, \mathcal{U}= \sum_{n=0}^{\infty}\, \frac{(-1)^n}{n!}\, \left[B, O \right]_n\, ,
\label{eq:conm_series}
\enq
which inevitably arises as one applies the Hadamard's lemma. As a result of this truncation, the calculation of any expectation value  $\la \Psi_{NG}\, |\, O\, |\, \Psi_{NG}\ra$ reduces to the computation of a finite number of Gaussian expectation values (calculability). In addition, the exponential nature of $\mathcal{U}$ asures an extensive volume dependence of observables such as the energy of the system. In other words, the non-Gaussianities captured by a trial wavefunctional of the form \eqref{eq:NG_wavefunc} persits in the thermodynamic limit. Finally,  being $\mathcal{U}$ unitary, the normalization of the state is preserved to the one imposed to the gaussian wavefunctional.

In this work, we build a post-Gaussian class of wavefunctionals through an operator $B$ of the form\footnote{We refer the reader to Appendix \ref{sec:appD} for a different proposal for $B$ and to references \cite{polley89, ritschel90, ibanez, Melgarejo:2019, Melgarejo:2020, Melgarejo:2021, Qian:2022} for examples of $B$ operators in different applications and contexts.}
\beq\begin{split}\label{eq:generator}
B&= -\alpha\int_{\vp,\vq_1,\vq_2,\vq_3} \, f(\vp,\vq_1,\vq_2,\vq_3)\, \frac{\delta}{\delta \phi(-\vp)}\, \bar{\phi}(\vq_1)\, \phi(\vq_2)\, \phi(\vq_3)\, \D^{D}(\vp + \vq_1 + \vq_2 +\vq_3)\, \\
&= -\alpha\int_{\vp,\vq_1,\vq_2,\vq_3} \, f(\vp,\vq_1,\vq_2,\vq_3)\, \pi(\vp)\, \bar{\phi}(\vq_1)\, \phi(\vq_2)\, \phi(\vq_3)\, \D^{D}(\vp + \vq_1 + \vq_2 +\vq_3)\, ,
\end{split}
\enq
with $\int_{\vp,\vq_1,\vq_2,\vq_3} \equiv \int \Dp \, \Dq_1\,  \Dq_2\, \Dq_3$.  Here, $\alpha$ is a  variational parameter that keeps track on the deviation of the wavefunctional and any observable from the Gaussian case. Following \cite{polley89, ritschel90, ibanez}, it is understood that an efficient truncation of the commutator series in \eqref{eq:conm_series} is such a one that terminates after the first non-trivial term when \eqref{eq:conm_series} is applied to the canonical fields of the theory. To this end, it is straightforward to see that the variational function  $f(\vp,\vq_1,\ldots,\vq_m)$, that is symmetric w.r.t. the exchange of $\vq_i$'s \footnote{As a variational function,  $f(\vp,\vq_1,\vq_2,\vq_3)$ must  obtained through energy minimization. The procedure to find optimal values for $f(\vp,\vq_1,\ldots,\vq_m)$ is given in Appendix \ref{sec:appC}.}, must be constrained to observe
\beq
\label{eq:constraint}
f(\vp,\vq_1,\ldots,\vq_m)
=0 
\ , 
\quad \vp=\vq_i\, ,
\qquad\text{and} \qquad
f(\vp,\vq_1,\ldots,\vq_m)\, f(\vq_i,\vk_1,\ldots,\vk_m) = 0
\, ,
\enq 
for $m=1,2,3$. These constraints force  that the action of $\mathcal{U}$ on the canonical field operators of the model \eqref{eq:hamilt}, $\phi(\vk),\, \bar{\phi}(\vk)$ and $\pi(\vk),\, \bar{\pi}(\vk)$ is given by
\beq\begin{split}\label{eq:nlct1}
\phi(\vk) & \to \phi(\vk) + \alpha\, \Phi(\vk)\, \\
\bar{\phi}(\vk) & \to \bar{\phi}(\vk) + \alpha\, \bar{\Phi}(\vk)\, 
\end{split}
\enq
with
\beq\begin{split}\label{eq:nlct2}
\Phi(\vk) &= -\int_{\vp,\vq,\vr}\, f(\vk,\vp,\vq,\vr)\, \bar{\phi}(\vp)\, \phi(\vq)\, \phi(\vr)\, \D^{D}(\vk-\vp-\vq-\vr)\, ,\\
\bar{\Phi}(\vk) & = 0\, ,
\end{split}
\enq

and
\beq\begin{split}\label{eq:nlct3}
\pi(\vk) & \to \pi(\vk)+ \alpha\, \Pi(\vk)\, , \\
\bar{\pi}(\vk) & \to \bar{\pi}(\vk) + \alpha\, \bar{\Pi}(\vk)\, ,
\end{split}
\enq
with
\beq\begin{split}\label{eq:nlct4}
\Pi(\vk) &= -2 \int_{\vp,\vq,\vr}\, f(\vp,\vq,\vk,\vr)\, \pi(\vp)\, \bar{\phi}(\vq)\, \phi(\vr)\, \D^{D}(\vk-\vp-\vq-\vr)\\
\bar{\Pi}(\vk) &=  \int_{\vp,\vq,\vr}\, f(\vp,\vq,\vk,\vr)\, \pi(\vp)\, \phi(\vq)\, \phi(\vr)\, \D^{D}(\vk-\vp-\vq-\vr)\, .
\end{split}
\enq

The quantities in \eqref{eq:nlct2} and \eqref{eq:nlct4} are nonlinear field functionals that shift the degrees of freedom of the canonical fields of the theory by a nonlinear polynomial function of other degrees of freedom. 

Being $\mathcal{U}$ unitary, one can check that Eqs. \eqref{eq:nlct1}-\eqref{eq:nlct4} ensure that 
\beq\begin{split}
\left[ \phi(\vk),\, \pi(\vk')\right]&\to \left[ \phi(\vk) +\alpha\, \Phi(\vk),\, \pi(\vk')+ \alpha\, \Pi(\vk')\right] = \D^{D}(\vk + \vk')\, \\
\left[ \bar{\phi}(\vk),\, \bar{\pi}(\vk')\right]&\to \left[ \bar{\phi}(\vk) + \alpha\, \bar{\Phi}(\vk),\, \bar{\pi}(\vk')+ \alpha \bar{\Pi}(\vk')\right] = \D^{D}(\vk + \vk')\, \\
\left[ \phi(\vk),\, \bar{\pi}(\vk')\right]&\to \left[ \phi(\vk) +\alpha\, \Phi(\vk),\, \bar{\pi}(\vk')+ \alpha\, \bar{\Pi}(\vk')\right] = 0\, ,
\end{split}
\enq
\emph{i.e}, the canonical commutation relations (CCR) still hold under the nonlinear transformed fields. For this reason, the  transformations above are known as nonlinear canonical transformations (NLCT). 

A remarkable property of the wavefunctionals \eqref{eq:NG_wavefunc} generated through NLCTs is that non-Gaussian corrections to Gaussian correlation functions can be obtained in terms of a finite number of Gaussian expectation values. Let us illustrate this with an example that is relevant when computing the expectation value of the Hamiltonian \eqref{eq:hamilt}. Under the NLCT implemented through $B$ in Eq. \eqref{eq:generator} we  have
 \beq\begin{split}\label{eq:int_terms}
 &\la\la \bar{\phi}(\vk_1)\bar{\phi}(\vk_2)\phi(\vk_3)\phi(\vk_4)\ra \ra  =  \la \bar{\phi}(\vk_1)\bar{\phi}(\vk_2)\phi(\vk_3)\phi(\vk_4)\ra 
 - \alpha \la \bar{\phi}(\vk_1)\bar{\phi}(\vk_2)\Phi(\vk_3)\phi(\vk_4)\ra\\
  &- \alpha \la \bar{\phi}(\vk_1)\bar{\phi}(\vk_2)\phi(\vk_3)\Phi(\vk_4)\ra 
   + \alpha^2 \la \bar{\phi}(\vk_1)\bar{\phi}(\vk_2)\Phi(\vk_3)\Phi(\vk_4)\ra\\
 &=\la \bar{\phi}(\vk_1)\bar{\phi}(\vk_2)\phi(\vk_3)\phi(\vk_4)\ra 
 - 2 \alpha \int f\, \la \bar{\phi}(\vk_1)\bar{\phi}(\vk_2)\phi(\vk_3)\bar{\phi}(\vp)\, \phi(\vq)\, \phi(\vr)\ra\\
 &+ \alpha^2 \int f^2\, \la \bar{\phi}(\vk_1)\bar{\phi}(\vk_2) \bar{\phi}(\vp_1)\, \phi(\vq_1)\, \phi(\vr_1)\, \bar{\phi}(\vp_2)\, \phi(\vq_2)\, \phi(\vr_2)\ra
 \end{split}
 \enq
where $\la \la \cdots \ra \ra$ refers to an evaluation w.r.t the non-Gaussian wavefunctional \eqref{eq:NG_wavefunc} and $\la \cdots\ra$ refers to a Gaussian expectation value w.r.t \eqref{eq:gauss_ansatz}. For reading convenience, we have schematically compressed the details of the arguments of the $f$-variational functions. It must be noted that the Gaussian expectation values appearing above contain an equal amount of fields $\phi$ as conjugate fields $\bar{\phi}$. As it will be shown, this is a convenient feature in order to variationally cover the structure of the vertex $V^{(4)}(\vk_1,\vk_2,\vk_3,\vk_4)$.

\subsection{Post-Gaussian improved ground state energy density}
\label{sec:ng_energy}
Now we are interested in evaluating $\mathcal{E}_{NG} \sim \la\la  H \ra\ra$ with $H$ in \eqref{eq:hamilt}. In order to this, apart from the four-point function stated above, we must compute
\beq\begin{split}
\la \la \bar{\pi}(\vk) \pi(-\vk)\ra \ra &= \la \bar{\pi}(\vk) \pi(-\vk)\ra + \alpha\, \la\bar{\pi}(\vk)\, \Pi(-\vk) \ra + \alpha\, \la \bar{\Pi}(\vk) \pi(-\vk)\ra + \alpha^2\la\bar{\Pi}(\vk) \Pi(-\vk) \ra\\
\la \la \bar{\phi}(\vk) \phi(-\vk)\ra \ra &= 
\la \bar{\phi}(\vk) \phi(-\vk)\ra + \alpha\, \la\bar{\phi}(\vk)\, \Phi(-\vk) \ra\, , 
\end{split}
\enq
where we used \eqref{eq:nlct1} and \eqref{eq:nlct4}. 

When taking into account the \emph{truncation} constraints \eqref{eq:constraint}
 one obtains the same result as for the Gaussian ansatz, that is
\beq\begin{split}\label{eq:ng2points}
\la \la \bar{\pi}(\vk) \pi(-\vk)\ra \ra &= \la \bar{\pi}(\vk) \pi(-\vk)\ra = \frac{1}{4}\, G^{-1}(\vk)\\
\la \la \bar{\phi}(\vk) \phi(-\vk)\ra \ra &= \la \bar{\phi}(\vk) \phi(-\vk)\ra = G(\vk)\, . 
\end{split}
\enq
Remarkably, while under dipole transformations, the nonlinear field shift in Eq \eqref{eq:nlct2} does not posses a clear transformation, the correlators above clearly state that the dipole one function still behaves as $G(\vk)\to G(\vk +\bd)$. 

With this, now we compute the dipole-dipole interaction term in the Hamiltonian \eqref{eq:hamilt}. Before that, we note that when evaluating the interaction term with the non-Gaussian ansatz, the $\mathcal{O}(\alpha^0)$ term  amounts to the Gaussian expectation value \footnote{We introduce the notation $\left[\Dk\right] \equiv \Dk_1 \Dk_2 \Dk_3 \Dk_4\, \delta^{D}(\vk_1 +\cdots \vk_4)$}
\beq\begin{split}
& \int \left[\Dk \right] \, V^{(4)}(\vk_1,\cdots \vk_4)\,  \la \la \bar{\phi}(\vk_1)\bar{\phi}(\vk_2)\phi(\vk_3)\phi(\vk_4)\ra \ra =\\
&=\int \left[\Dk \right] \, V^{(4)}(\vk_1,\cdots \vk_4)\,   \la \bar{\phi}(\vk_1)\bar{\phi}(\vk_2)\phi(\vk_3)\phi(\vk_4)\ra + \mathcal{O}(\alpha, \alpha^2)
=\int \Dk\, G(\vk)\, \Sigma(\vk) + \mathcal{O}(\alpha, \alpha^2)\, ,
\end{split}
\enq
with $\mathcal{O}(\alpha, \alpha^2)$ refering to non-Gaussian corrections. To be more explicit, denoting the Gaussian energy expectation value by $\mathcal{E}_G$, one may write
\beq\label{eq:ng_energy_vev}
\mathcal{E}_{NG}=\mathcal{E}_G + \Delta\, \mathcal{E}^{(1)} + \Delta\, \mathcal{E}^{(2)}\, ,
\enq
where $\Delta\, \mathcal{E}^{(1)}$ and $\Delta\, \mathcal{E}^{(2)}$ refer to $\mathcal{O}(\alpha)$ and $\mathcal{O}(\alpha^2)$ non-Gaussian corrections respectively. Interestingly, these correction terms are completely related to the connected part of the four-point function $\la \la \bar{\phi}(\vk_1)\bar{\phi}(\vk_2)\phi(\vk_3)\phi(\vk_4)\ra \ra$. That is, substracting the disconnected terms in Eq. \eqref{eq:int_terms} one obtains
\beq\begin{split}\label{eq:fourpoint}
\la \la \bar{\phi}(\vk_1)\bar{\phi}(\vk_2)\phi(\vk_3)\phi(\vk_4)\ra \ra_{\rm conn}&= -2 \alpha \int f\, \la \bar{\phi}(\vk_1)\bar{\phi}(\vk_2)\phi(\vk_3)\bar{\phi}(\vp)\, \phi(\vq)\, \phi(\vr)\ra\\
 & + \alpha^2 \int f^2\, \la \bar{\phi}(\vk_1)\bar{\phi}(\vk_2) \bar{\phi}(\vp_1)\, \phi(\vq_1)\, \phi(\vr_1)\, \bar{\phi}(\vp_2)\, \phi(\vq_2)\, \phi(\vr_2)\ra\, .
 \end{split}
\enq
In words, the connected part of the four point function completely determines the  the non-Gaussian corrections to the ground state energy or an improved version of the effective potential in \eqref{eq:gep}. 

\paragraph*{$\Delta\, \mathcal{E}^{(1)}$ correction.} This term corresponds to the evaluation of $\chi_1^{(1)} + \chi_2^{(1)}$ with,
\beq\begin{split}
\chi_1^{(1)}&=\int \left[\Dk \right] \, V^{(4)}(\vk_1,\cdots \vk_4)\,  \la \bar{\phi}(\vk_1)\bar{\phi}(\vk_2)\Phi(\vk_3)\phi(\vk_4)\ra \, \\
\chi_2^{(1)}&=\int \left[\Dk \right] \, V^{(4)}(\vk_1,\cdots \vk_4)\,  \la \bar{\phi}(\vk_1)\bar{\phi}(\vk_2)\phi(\vk_3)\Phi(\vk_4)\ra\, .
\end{split}
\enq
By symmetry, $\chi_1^{(1)} + \chi_2^{(1)}\equiv \chi^{(1)}$ so we focus on the first one, which  explicitly reads
\beq\begin{split}
\chi^{(1)} = -\alpha \int_{\vp,\vq,\vr} \left[\Dk \right]\,  V^{(4)}(\vk_i)\,  \tilde{f}(\vk_3,\vp,\vq,\vr)\, \la \bar{\phi}(\vk_1)\bar{\phi}(\vk_2)\bar{\phi}(\vp)\phi(\vq)\phi(\vr)\phi(\vk_4)\ra\, 
\end{split}
\enq 
with $\tilde{f}(\vk_3,\vp,\vq,\vr)\equiv f(\vk_3,\vp,\vq,\vr)\, \D^{D}(\vk_3-\vp-\vq-\vr)$. 
Using the truncation constraints \eqref{eq:constraint} and Wick's theorem, the result reads as
\beq\begin{split}
\chi^{(1)}&=-\alpha \int_{\vp,\vq,\vr}\, \left[\Gamma^{(4)}_1(\vp,\vq,\vr) + \Gamma^{(4)}_1(-\vp,\vq,\vr)\right]\, G(\vp)\, G(\vq)\, G(\vr)\, ,
\end{split}
\enq
with
\beq\label{eq:Gamma1}
\Gamma^{(4)}_1(\vp,\vq,\vr) = c(\vp,\vq,\vr)\, V^{(4)}(-\vr,-\vq,\vp+\vq+\vr,-\vp)\,
\enq
where we have used the compact notation
\beq
c(\vp,\vq,\vr)\equiv f(\vp+\vq+\vr,\vp,\vq,\vr)\, .
\enq
Therefore, the $\mathcal{O}(\alpha)$-contribution to the post-Gaussian improved energy expectation value $\Delta\, \mathcal{E}^{(1)}$ can be written as
\beq\label{eq:alpha1_correction}
\Delta\, \mathcal{E}^{(1)} = 2\, \chi^{(1)} = -2\, \alpha \int_{\vp,\vq,\vr}\, \left[\Gamma^{(4)}_1(\vp,\vq,\vr) + \Gamma^{(4)}_1(-\vp,\vq,\vr)\right]\, G(\vp)\, G(\vq)\, G(\vr)\, ,
\enq
It is worth to remark that in \eqref{eq:alpha1_correction}, the "variationally dressed vertex" $\Gamma^{(4)}_1(\vp,\vq,\vr)$,  captures a range of values of \eqref{eq:vertex} larger than the forward scattering limit of the Gaussian term. In other words, $\Delta\, \mathcal{E}^{(1)}$  grasps features of the tensor part of the dipole-dipole interaction that are inaccessible to the Gaussian approach.

\paragraph*{$\Delta\, \mathcal{E}^{(2)}$ correction.} This term corresponds to the evaluation of
\beq\begin{split}
\chi^{(2)}&=\alpha^{2}\int \left[\Dk\right]\left[D_1\right]\left[D_2\right]\, V^{(4)}(\vk_i)\, \tilde{f}(\vk_3,\vp_1,\vq_1,\vr_1)\, \tilde{f}(\vk_4,\vp_2,\vq_2,\vr_2)\\
&  \times \la \bar{\phi}(\vk_1)\bar{\phi}(\vk_2)\bar{\phi}(\vp_1)\phi(\vq_1)\phi(r_1)\bar{\phi}(\vp_2)\phi(\vq_2)\phi(\vr_2)\ra\\
\end{split}
\enq
with $\left[D_i\right]\equiv \left[D\right]_{p_i,q_i,r_i}\, i=1,2$. After a lengthy albeit straightforward calculation, the result reads as
\beq\label{eq:alpha2_terms}
\Delta\, \mathcal{E}^{(2)}=\chi^{(2)}=4\alpha^{2}\int_{\substack{\vp_1,\vp_2\\ \vq_1,\vq_2}}\, \Gamma^{(4)}_2(\vp_1,\vp_2,\vq_1,\vq_2) \,  G(\vp_1) G(\vp_2) G(\vq_1) G(\vq_2) \,  .
\enq
with
\beq\label{eq:Gamma2}
\Gamma^{(4)}_2(\vp_1,\vp_2,\vq_1,\vq_2) = c(\vp_1,-\vp_2,\vq_1)\, c(\vp_2,-\vp_1,\vq_2)\, V^{(4)}(-\vq_2,-\vq_1,\vp_1-\vp_2 + \vq_1, \vp_2-\vp_1+ \vq_2)\, .
\enq
As before, the variationally dressed vertex $\Gamma^{(4)}_2(\vp,\vq,\vr,\vz)$ covers a wider range of values of the vertex interaction \eqref{eq:vertex}  than the forward scattering Gaussian limit, thus capturing the $\lambda_T$ part of the dipole-dipole interaction.

The post-Gaussian improved energy density $\mathcal{E}_{NG}$ in Eqs \eqref{eq:ng_energy_vev}, \eqref{eq:alpha1_correction} and \eqref{eq:alpha2_terms} as said above, is builded in terms of a field transformations that shift some degrees of freedom of the canonical fields through a nonlinear polynomial function of other degrees of freedom. This is implemented by the structure of the variational parameters $c(\vp,\vq,\vr)$. Concretely, in Appendix \ref{sec:appC} we adopt a functional structure for $c(\vp,\vq,\vr)$ such that the IR modes of $\phi$ are nonlinearly shifted by its UV modes. In this sense, it is expected that an optimized set of $c(\vp,\vq,\vr)$ parameters can conform, in a variational sense, the integration of the UV modes to the symmetries of the model \eqref{eq:fracton_lagrangian}.

\begin{figure*}
    \centering
    \includegraphics[width = 0.97\linewidth]{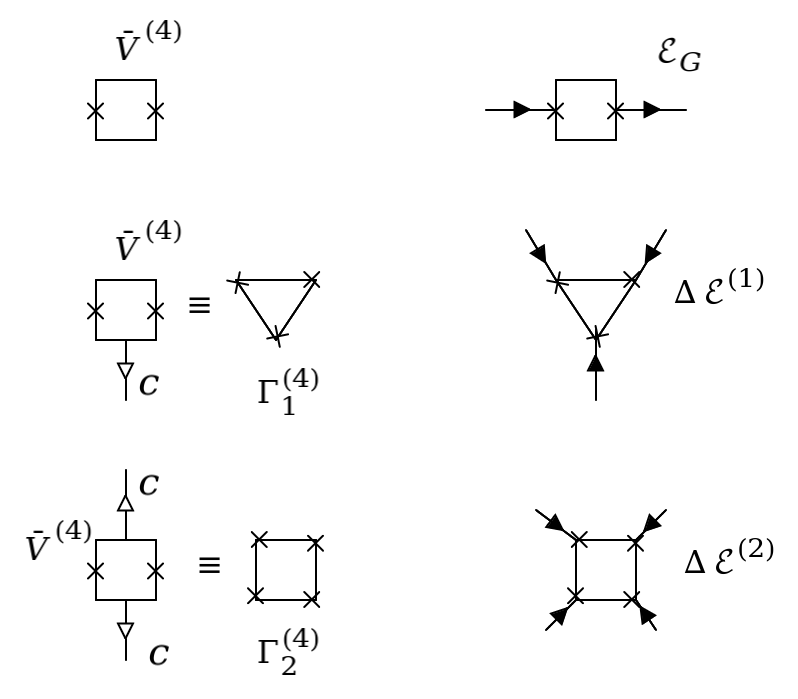}
    \caption{Diagramatic representation of terms in \eqref{eq:ng_energy_vev}. Up: The dipole-dipole interaction part of the Gaussian term (\emph{right}) involves two propagators $G(\vk)$ (black arrow) attached to the vertex $V^{(4)}$ (\emph{left}) with two insertion points (cross). Middle: Diagramatic representation of \eqref{eq:Gamma1} (\emph{left}). The variational parameter $c(\vp,\vq,\vr)$ (white arrow) combines with $V^{(4)}$ to build the "3-legged" variational vertex $\Gamma^{(4)}_1$ and the correction to energy density $\Delta\, \mathcal{E}_1$ Eq \eqref{eq:alpha1_correction}. Down: Diagramatic representation of \eqref{eq:Gamma2} (\emph{left}). Two variational parameters $c(\vp,\vq,\vr)$ (white arrows) combine with $V^{(4)}$ to build the "4-legged" variational vertex $\Gamma^{(4)}_2$ and the correction to energy density $\Delta\, \mathcal{E}_2$ Eq. \eqref{eq:alpha2_terms}.}
    \label{fig:feynman}
\end{figure*}

This integration helps to build the "effective variational vertices" $\Gamma^{(4)}_1(\vp,\vq,\vr)$ and $\Gamma^{(4)}_2(\vp,\vq,\vr,\vz)$ involving three and four propagators $G(\vk)$ respectively. In this sense, noting that (in the forward scattering limit) $V^{(4)}$ can be understood as a vertex involving two propagators $G(\vk)$, $\Gamma^{(4)}_1$ in \eqref{eq:Gamma1} is the combination  $c\, -\, V^{(4)}$ and $\Gamma^{(4)}_2$ in \eqref{eq:Gamma2} amounts to $c\, -\, V^{(4)}\, -\, c\,$. This suggests the interpretation that each $c(\vp,\vq,\vr)$ creates a new "insertion" point into the variational vertex to which a new propagator can be attached. This is pictorially shown in Fig \ref{fig:feynman} where the different terms in Eq. \eqref{eq:ng_energy_vev} are presented diagramatically.

It is important to remark at this point that general equations for the optimal values of the variational parameters  $G(\vk), \Sigma(\vk)$ and $c(\vp,\vq,\vr)$ can be obtained, given a fixed $\alpha$\footnote{Note that $\alpha$ is nothing more that a normalization value for the parameters $c(\vp,\vq,\vr)$ which we use as a tracking parameter. }, by deriving $\mathcal{E}_{NG}$ w.r.t. them and then equating to zero. This yields a set of non-linear coupled equations that must be self-consistently solved. Finding analytical  expressions for this solutions is difficult. In this regard, our aim here is to provide expressions that explicitly show the relation between the variational parameters and the coupling constants of the model under consideration. With this aim, from here in advance we choose $\alpha$ in such a way that \beq\label{eq:ng_scaling_limit}
\alpha\, \lambda_S\ll 1\, \quad \alpha\, \lambda_T\ll 1\, ,
\enq
with the coupling constants $\lambda$'s not necessarily small. With this choice, the optimization equations greatly simplify and \emph{i)} $G(\vk)$ and $\Sigma(\vk)$ can be reduced to the Gaussian solution Eq. \eqref{eq:Sigma_condensate} and \emph{ii)} the optimization of the variational parameters $c(\vp,\vq,\vr)$ is decoupled from them and can be carried over the non-Gaussian correction $\Delta\mathcal{E}^{(1)} + \Delta\mathcal{E}^{(2)}$. The simplification occurs as a consequence of the structure of the optimization equations (discussed in Appendix \ref{sec:appC}) and is never due to any additional assumptions.  It is clear that if one chooses $\alpha$ not fulfilling \eqref{eq:ng_scaling_limit} for solving the equations, $G(\vk), \Sigma(\vk)$ have not the Gaussian structure anymore. In the limit posed by \eqref{eq:ng_scaling_limit}, it is obvious that the main contribution for a post-Gaussian improvement of the ground state energy of the theory in \eqref{eq:fracton_lagrangian} is given by $\Delta\, \mathcal{E}^{(1)}$. 

\subsection{Renormalization and $\beta$-functions}
\label{sec:ng_renormalization}
In this subsection, we obtain an improved effective potential with the aim of studying the RG flow of the model \eqref{eq:fracton_lagrangian} through the $\beta$-functions for the dipole-dipole interaction. 

The renormalization of post-Gaussian variational calculations is rather involved. In \cite{ritschel:91} it was showed that contributions generated by the NLCT-wavefunctionals, which are related to 1-particle irreducible diagrams not taken into account by the Gaussian approximation, require new counterterms that in principle could be  determined systematically by taking the derivatives of the effective potential. Nevertheless, this approach is preconditioned by an analytic optimization of the energy expectation value, that, in general, is not feasible (see Appendix \ref{sec:appC}). 

Fortunately, it is possible to make advance by using simplified functional forms for the variational parameters $c(\vp,\vq,\vr)$. This approach can be taken as a compromise route to gain further understanding of the problem \cite{polley89, ritschel90, ritschel:91, ibanez}; while it leads to a sub-optimal approximation, a particular ansatz for the NLCT $c(\vp,\vq,\vr)$, motivated by the optimization of a part of the energy expectation, amounts to a renormalized post-Gaussian effective potential (see Appendix \ref{sec:appC}). 

The following discussion thus takes for granted that the (sub-)optimal values of the variational parameters of the ansatz have been found.

\paragraph{NLCT-Improved Effective Potential.} Here we consider the the non-Gaussian corrections to the Gaussian effective potential in \eqref{eq:gep} generated by the wavefunctional 
\beq\label{eq:NG_wavefunc_sigma}
\Psi_{NG}[\phi,\,  \bar{\phi}] =  \mathcal{U}\, \Psi_{SG}[\phi,\,  \bar{\phi}]= \exp(B)\, \Psi_{SG}[\phi,\,  \bar{\phi}]\, ,
\enq
with $\Psi_{SG}$ defined in \eqref{eq:SG_Gaussian}. These corrections are given by
\beq
\Delta\, \mathbb{V}_{\rm eff}[\sigma] = \Delta \mathbb{V}^{0}_{\rm eff}[\sigma] +
\sigma^{2}\, \Delta \mathbb{V}^{2}_{\rm eff}[\sigma] +  \sigma^{4}\, \Delta \mathbb{V}^{4}_{\rm eff}[\sigma]\, ,
\enq
with
\beq\begin{split}
\Delta \mathbb{V}^{0}_{\rm eff}[\sigma]&=\left(-2\alpha \mathbb{V}^{(1)}_{0}[\sigma] + 4 \alpha^2\, \mathbb{V}^{(2)}_{0}[\sigma]\right)\, , \\
\Delta \mathbb{V}^{2}_{\rm eff}[\sigma]&=\left(-2\alpha \mathbb{V}^{(1)}_{2}[\sigma] + 4 \alpha^2\, \mathbb{V}^{(2)}_{2}[\sigma]\right)\, ,\\
\Delta \mathbb{V}^{4}_{\rm eff}[\sigma]&=4 \alpha^2\,  \mathbb{V}^{(2)}_{4}[\sigma]\, ,
\end{split}
\enq

and
\beq\begin{split}
\mathbb{V}^{(1)}_{0}[\sigma]&= \int_{\vp,\vq,\vr}\, \left[\Gamma^{(4)}_1(\vp,\vq,\vr) + \Gamma^{(4)}_1(-\vp,\vq,\vr)\right]\, G(\vp)\, G(\vq)\, G(\vr)\, \\
\mathbb{V}^{(2)}_{0}[\sigma]&=\int_{\vp,\vq,\vr,\vz}\, \Gamma^{(4)}_2(\vp,\vq,\vr,\vz)\, G(\vp) G(\vq) G(\vr) G(\vz) \, \\
\mathbb{V}^{(1)}_{2}[\sigma] &= \int_{\vp,\vq}\,\left[\Gamma^{(4)}_1(\vp,\vq,0) + \Gamma^{(4)}_1(-\vp,\vq,0)\right]\, G(\vp) G(\vq)+ \, \int_{\vq,\vr}\,\Gamma^{(4)}_1(0,\vq,\vr)\, G(\vq) G(\vr)\, ,\\
\mathbb{V}^{(2)}_{2}[\sigma] &= \int_{\vp_1,\vp_2,\vq_1} \Gamma^{(4)}_2(\vp_1,\vp_2,\vq_1,0) G(\vp_1)G(\vp_2)G(\vq_1) + \int_{\vp_1,\vp_2,\vq_2} \Gamma^{(4)}_2(\vp_1,\vp_2,0,\vq_2) G(\vp_1)G(\vp_2)G(\vq_2)\\
& + \int_{\vp_1,\vq_1,\vq_2} \Gamma^{(4)}_2(\vp_1,0,\vq_1,\vq_2) G(\vp_1)G(\vq_1)G(\vq_2) +  \int_{\vp_2,\vq_1,\vq_2} \Gamma^{(4)}_2(0,\vp_2,\vq_1,\vq_) G(\vp_2)G(\vq_1)G(\vq_2)\, , \\
\mathbb{V}^{(2)}_{4}[\sigma] &= \int_{\vp,\vq}\, \Gamma^{(4)}_2(\vp,\vq,0,0) G(\vp)\, G(\vq)\, .
\end{split}
\enq

According to the definitions in \eqref{eq:Gamma1} and \eqref{eq:Gamma2}, we observe that while $\Delta \mathbb{V}^{0}_{\rm eff}[\sigma]$ and $\Delta \mathbb{V}^{2}_{\rm eff}[\sigma]$ capture both the $T$-part and the $S$-part of \eqref{eq:vertex}, the term $\Delta \mathbb{V}^{4}_{\rm eff}[\sigma]$ only captures the $S$-part. It is thus convenient for our interests to split  $\mathbb{V}_{\rm eff}$ into a $T$-term and $S$-term. To this end, we decompose \eqref{eq:vertex} as
\beq\label{eq:vertex_decomposed}
V^{(4)}(\vk_{12},\vk_{34}) \equiv \frac{\lambda_T}{4}\, \widetilde{V}_T^{(4)}(\vk_{12},\vk_{34})+  \frac{\lambda_S}{4}\, \widehat{V}_S^{(4)}(\vk_{12},\vk_{34})\, ,
\enq
with
\beq\begin{split}
\widetilde{V}_T^{(4)}(\vk_{12},\vk_{34})&=\left[ \left(\vk_{12}\cdot \vk_{34}\right)^2 - |\vk_{12}|^2 |\vk_{23}|^2\right]\, ,\\
\widehat{V}_S^{(4)}(\vk_{12},\vk_{34})&=\left[ \left(|\vk_{12}|^2 + 2\bar{\gamma}\right)\left(|\vk_{34}|^2 + 2\gamma\right)\right]\, .
\end{split}
\enq

After this splitting, one can write the post-Gaussian improved effective potential as
\beq\label{eq:ng_ef_pot}
\mathbb{V}_{\rm eff}[\sigma] = \mathbb{V}^{S}_{\rm eff}[\sigma] + \mathbb{V}^{T}_{\rm eff}[\sigma]\, 
\enq
with
\beq
\mathbb{V}^{S}_{\rm eff}[\sigma]= \mathbb{V}^{G}_{\rm eff}[\sigma] + \Delta \mathbb{V}^{0\, (S)}_{\rm eff}[\sigma] + \sigma^2\, \Delta \mathbb{V}^{2\, (S)}_{\rm eff}[\sigma] + \sigma^4\, \Delta \mathbb{V}^{4\, (S)}_{\rm eff}[\sigma]\, , 
\enq
and
\beq
\mathbb{V}^{T}_{\rm eff}[\sigma]=  \Delta \mathbb{V}^{0\, (T)}_{\rm eff}[\sigma] + \sigma^2\, \Delta \mathbb{V}^{2\, (T)}_{\rm eff}[\sigma] \, .
\enq
Here, $\mathbb{V}^{G}_{\rm eff}[\sigma]$ refers to the Gaussian effective potential in \eqref{eq:gep} and the $T$ and $S$ labels refer to the vertex splitting in \eqref{eq:vertex_decomposed}.

\paragraph{RG flow and $\beta$-functions.} In order to obtain the running of the coupling constants with the energy scale, following \cite{Coleman:1973} as before, we define the renormalized couplings as
\beq\label{eq:ng-running-couplings}
\widehat{\lambda}_{M} = \frac{1}{2}\, \frac{d^{2}\, \mathbb{V}^{S}_{\rm eff}[\sigma]}{d (\sigma^{2})^2}\Bigg|_{\sigma=M}\, ,\qquad 
\widetilde{\lambda}_{M} = \frac{1}{2}\, \frac{d^{2}\, \mathbb{V}^{T}_{\rm eff}[\sigma]}{d (\sigma^{2})^2}\Bigg|_{\sigma=M}\, .
\enq
At this point, we note the following: by working in the scaling limit where $\alpha\, \lambda_{T,\, S} \ll 1$, it is reasonable to consider that a good approximation for the renormalized $S$-coupling and its RG-flow $\beta$-function is given by the Gaussian term, that is to say, $\widehat{\lambda}_{M}$ is given by Eq. \eqref{eq:renormalized_coupling} and 
\beq
\beta(\widehat{\lambda}_{M}) = 3\, \widehat{\lambda}_{M}^3\, \Lambda^{S}(M) + \mathcal{O}(\alpha)\, ,
\enq 
where $\Lambda^{S}(M)$ is the one appearing in \eqref{eq:beta_func_M}. It is obvious that the non-Gaussian terms provide $\alpha$-corrections to the $S$-interaction that might be interesting to compute. However, we foresee that a bigger interest relies on seeing how the non-Gaussian terms in \eqref{eq:ng_ef_pot} provide a non-vanishing $\beta$-function for the $T$-interaction part between dipoles. Therefore, here we will focus on $\mathbb{V}^{T}_{\rm eff}[\sigma]$ that, conveniently rearranged, reads as
\beq\label{eq:t-ngep}
\mathbb{V}^{T}_{\rm eff}[\sigma]=  -2\alpha \left(\mathbb{V}^{(1)T}_{0}[\sigma] +\sigma^2\, \mathbb{V}^{(1)T}_{2}[\sigma]\right) + 4\alpha^2\, \left(\mathbb{V}^{(2)T}_{0}[\sigma] +\sigma^2\, \mathbb{V}^{(2)T}_{2}[\sigma]\right)\, .
\enq

In addition, we note that when plugging \eqref{eq:t-ngep} into \eqref{eq:ng-running-couplings}, in the limit \eqref{eq:ng_scaling_limit}, the contributions coming from $\mathcal{O}(\alpha^2)$ terms in \eqref{eq:t-ngep} can be effectively discarded and keep  only considering the $\mathcal{O}(\alpha)$ terms in way such that,
\beq\label{eq:t-ngep_approx}
\mathbb{V}^{T}_{\rm eff}[\sigma]\approx  -2\alpha \left(\mathbb{V}^{(1)T}_{0}[\sigma] +\sigma^2\, \mathbb{V}^{(1)T}_{2}[\sigma]\right) \, ,
\enq
with
\beq\begin{split}
\mathbb{V}^{(1)T}_{0}[\sigma]&= \int_{\vp,\vq,\vr}\, \left[\Gamma^{(4) T}_1(\vp,\vq,\vr) + \Gamma^{(4) T}_1(-\vp,\vq,\vr)\right]\, G(\vp)\, G(\vq)\, G(\vr)\, \\
\mathbb{V}^{(1)T}_{2}[\sigma] &= \int_{\vp,\vq}\,\left[\Gamma^{(4) T}_1(\vp,\vq,0) + \Gamma^{(4)\, T}_1(-\vp,\vq,0)\right]\, G(\vp) G(\vq)+ \, \int_{\vq,\vr}\,\Gamma^{(4) T}_1(0,\vq,\vr)\, G(\vq) G(\vr)\, ,
\end{split}
\enq
and
\beq
\Gamma^{(4) T}_1(\vp,\vq,\vr) = \frac{\lambda_T}{4}\, c(\vp,\vq,\vr)\, \widetilde{V}_T^{(4)}(\vq-\vr,2\vp +\vq+\vr)\, .
\enq

With this we write,
\beq\label{eq:lambdaT-M}
\widetilde{\lambda}_{M} =  -2\alpha\, \left( \mathbb{I}^{(1)}_{0}(M) + \mathbb{I}^{(1)}_{2}(M)\right)\, ,
\enq
where
\beq\label{eq:lambdaT-terms}
\mathbb{I}^{(1)}_{0}(M)=\frac{1}{2}\, \frac{d^{2}\, \mathbb{V}^{(1)T}_{0}[\sigma]}{d (\sigma^{2})^2}\Bigg|_{\sigma=M}\, ,\quad  \mathbb{I}^{(1)}_{2}(M)=\frac{1}{2}\, \frac{d^{2}}{d (\sigma^{2})^2}\,  \left(\sigma^2\, \mathbb{V}^{(1)T}_{2}[\sigma]\right)\Bigg|_{\sigma=M}\, .
\enq

Let us first focus on $\mathbb{I}^{(1)}_{2}(M)$. Specifically, let us note the result of the second derivative w.r.t $\sigma^2$ of the arguments inside the momentum integrals yields
\beq\begin{split}\label{eq:relevant_ng_term}
&-2\alpha\, \left(\frac{1}{2}\, \frac{d^{2}}{d (\sigma^{2})^2}\, \sigma^2\, G(\vk) G(\vk')\right)\Bigg|_{\sigma=M} = 8\alpha \Bigg[\Bigg(V^{(S)}_{\vk'} G(\vk) G(\vk')^3 + (\vk \leftrightarrow \vk')\Bigg)\\
&-2M^2 \Bigg( 3 (V^{(S)}_{\vk'})^2 G(\vk) G(\vk')^5 + (\vk \leftrightarrow \vk') +2 V^{(S)}_{\vk} V^{(S)}_{\vk'} G(\vk)^3 G(\vk')^3\Bigg)\Bigg]\, ,
\end{split}
\enq
with $V^{(S)}_{\vk}\equiv\bar{V}^{(4)}(\vk,0)$. 

At this point we remark that terms with higher number of "propagators" $G(\vk)$ and "vertices" $V^{(S)}_{\vk}$ correspond to higher loop terms in the dipole-dipole interaction captured by the non-Gaussian ansatz. Therefore, it is sensible that we keep in our calculation, only those terms with the lowest number of them. In this sense, the terms multiplied by the arbitrary substraction point $M$ involve a high number of $G(\vk)$'s, and we will assume that they can be safely discarded w.r.t the other terms. Under the same assumption, $\mathbb{I}^{(1)}_{0}(M)$, which contain terms with two "vertices" $V^{(S)}_{\vk}$ and seven "propagators" $G(\vk)$, may be discarded as well. With this, one can approximate
\beq\label{eq:lambdaT_M}
\widetilde{\lambda}_{M} \approx \frac{\alpha}{2}\, (\lambda_T\, \lambda_S)\, I^{T}(M)\, ,
\enq
where $\lambda_S$ and $\lambda_T$ are the bare parameters in \eqref{eq:fracton_lagrangian} and
\beq\begin{split}
I^{T}(M) &= \Bigg[\int_{\vp,\vq}\, \left[c(\vp,\vq,0)\, \widetilde{V}_T^{(4)}(\vq,2\, \vp +q) + (\vp \leftrightarrow -\vp)\right] \left(V^{(S)}_{\vq}\, G(\vp) G(\vq)^3 + (\vp \leftrightarrow \vq)\right)\\
&+ \int_{\vq,\vr}\, c(0,\vq,\vr)\, \widetilde{V}_T^{(4)}(\vq-\vr,\vq + \vr) \left(V^{(S)}_{\vq}\,  G(\vr) G(\vq)^3 + (\vq \leftrightarrow \vr)\right)
\Bigg]\, .
\end{split}
\enq
A renormalized $T$-coupling can be defined by taking the substraction point to be $M=0$, as
\beq
\widetilde{\lambda}_{R} = \frac{\alpha}{2}\, (\lambda_T\, \lambda_S)\, I^{T}(0)\, , 
\enq
that, interestingly, depends on both bare parameters of the model. It is remarkable also that it is possible to define this renormalized coupling only for $\alpha \neq 0$.

Finally, we compute the RG flow of $\widetilde{\lambda}_{M}$, which after a lenghty albeit straightforward computation yields
\beq\label{eq:beta_func_T}
\beta(\widetilde{\lambda}_M)= M\, \frac{\partial \widetilde{\lambda}_{M}}{\partial M}=-\alpha\, \lambda_T\, \lambda_S^2\, \Lambda^T(M)\, ,
\enq
with
\beq
\Lambda^T(M) = M^2\, \mathcal{I}^T(M)\, ,
\enq
and
\beq\begin{split}
\mathcal{I}^T(M) &= \int_{\vp,\vq}\, \Bigg( h^1_1(\vp,\vq,0)\, G(\vp)G(\vq)^3 \left[3 \hat{V}_{\vq}^{(S)} G(\vq)^2 +\hat{V}_{\vp}^{(S)} G(\vp)^2\right]\\
&+ h^1_2(\vp,\vq,0)\, G(\vp)^3G(\vq) \left[3 \hat{V}_{\vp}^{(S)} G(\vp)^2 +\hat{V}_{\vq}^{(S)} G(\vq)^2\right]
\Bigg)\\
& + \int_{\vq,\vr}\, \Bigg( h^2_1(0,\vq,\vr)\, G(\vr)G(\vq)^3 \left[3 \hat{V}_{\vq}^{(S)} G(\vq)^2 +\hat{V}_{\vr}^{(S)} G(\vr)^2\right]\\
&+ h^2_2(0,\vq,\vr)\, G(\vr)^3G(\vq) \left[3 \hat{V}_{\vr}^{(S)} G(\vr)^2 +\hat{V}_{\vq}^{(S)} G(\vq)^2\right]
\Bigg)
\end{split}
\enq
where we use the compact notation $\hat{V}_{\vk}^{(S)}\equiv ||\vk|^2 + 2\gamma|^2$ and
\beq\begin{split}
h^1_1(\vp,\vq,0) &= \hat{V}_{\vq}^{(S)}\, \left[c(\vp,\vq,0)\, \widetilde{V}_T^{(4)}(\vq,2\, \vp +\vq) + (\vp \leftrightarrow -\vp)\right]\, ,\\
h^1_2(\vp,\vq,0) &= \hat{V}_{\vp}^{(S)}\, \left[c(\vp,\vq,0)\,  \widetilde{V}_T^{(4)}(\vq,2\, \vp +\vq) + (\vp \leftrightarrow -\vp)\right]\, ,\\
h^2_1(0,\vq,\vr) &= \hat{V}_{\vq}^{(S)}\, c(0,\vq,\vr)\,  \widetilde{V}_T^{(4)}(\vq-\vr,\vq+\vr)\, ,\\
h^2_2(0,\vq,\vr) &= \hat{V}_{\vr}^{(S)}\, c(0,\vq,\vr)\,  \widetilde{V}_T^{(4)}(\vq-\vr,\vq+\vr)\, .
\end{split}
\enq

Let us summarize our results as
\beq\begin{split}\label{eq:beta_func_ng}
\widehat{\beta}=\beta(\widehat{\lambda}_M) &=  3\, \lambda_S^3\, \Lambda^{S}(M) + \mathcal{O}(\alpha)\, ,\\
\widetilde{\beta}=\beta(\widetilde{\lambda}_M)&=-\alpha\, \lambda_T\, \lambda_S^2\, \Lambda^T(M) + \mathcal{O}(\alpha^2) \, ,
\end{split}
\enq
where we note that $\Lambda^{S}(M)$  in the first equation is a rescaled version of the one appearing in \eqref{eq:beta_func_M}. Eqs \eqref{eq:beta_func_ng} represent the main results of this work for several reasons:
\begin{enumerate}
\item First, they show that our post-Gaussian ansatz, by providing (-a variational approximation of-) the connected part of the four point function (a.k.a dipole two point funcion) is able to capture \emph{i)} the transverse component of the dipole-dipole interaction and consequently, \emph{ii)} a nonvanishing $\beta$-function for this part of the interaction. This is something that cannot be derived from a pure Gaussian ansatz or large $N$ techniques.

\item As it was commented in Section \ref{sec:model}, the Hamiltonian \eqref{eq:hamilt} is bounded from below for $\lambda_T\geq 0$ and $\lambda_T + D\,  \lambda_S \geq 0$ \cite{Bidussi:2021}. Thus, our non-perturbative calculation of the $\beta$-functions shows an intringuing feature. Noting that, being $\Lambda^{T}(M)$  a priori, a non negative function of $M^2$, the result in \eqref{eq:beta_func_ng} suggests that, with respect to the transverse part of the dipole-dipole interaction, the model is \emph{asymptotically free}, that is, the renormalized coupling $\widetilde{\lambda}_{M}$ decreases as we move into the UV.  This is a highly non-trivial result. This kind of dipole-\emph{asymptotic freedom} suggests that symmetries that manifest only due to the strong coupling of dipoles, may arise at low energies from less symmetric, in the dipole invariance sense, UV models. Of course, it would be customary to check, by carrying out explicit optimizations of the ansatz, if $ \Lambda^{T}(M)$ is always positive or posses zero crossings that would imply more involved behaviours.

\item Regarding the last point, there is an interesting limit for the model in Eq. \eqref{eq:fracton_lagrangian} that can be reached by taking $\lambda_S\to 0^{+}$ while holding $\lambda_S\, \lambda_T = \lambda$ fixed. In this limit, the renormalized coupligs are well defined and read as
\beq
\widehat{\lambda}_R \to 0\, \quad  {\rm and}\quad \widetilde{\lambda}_R \to \frac{\alpha}{2}\, \lambda\, I^{T}(0)\, .
\enq  
In addition, both $\beta$-functions vanish, with 
\beq
\widehat{\beta} \sim \lambda_S^3 \to 0\, ,
\enq
as we are in the non-interaction point for the $S$-part of the dipole-dipole interaction, and \beq
\widetilde{\beta} \sim \lambda\, \lambda_S \to 0\, .
\enq 
The latter implies that the tensor part of the dipole-dipole coupling does not flow for any arbitrary initial bare value of $\lambda_T$. From this point of view, in this limit, the model amounts to a well-defined quantum field theory with fractonic behaviour, a result that seems to be consistent with those in \cite{Distler:2021} and \cite{Grosvenor:2022}. The RG flow diagram plotted in Fig. \ref{fig:beta_streams}, shows the limit commented above in the region where the strenghth of the flow is close to zero (dark region). 

\begin{figure*}
    \centering
    \includegraphics[width = 0.57\linewidth]{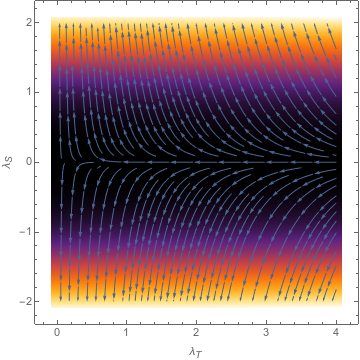}
    \caption{The RG flow implemented by Eqs \eqref{eq:beta_func_ng}. The figure plots the vector field $(\tilde{\beta}, \hat{\beta})\vert_{M=0}$ and it has been generated assuming that $\Lambda^T(0) \sim \Lambda^S(0)$. The arrows indicate the direction of the flow and the background color refers to the strengh of the flow, from weak or zero (dark) to a bigger values (orange). In the dark central region the model does not flows and is a well defined QFT. The rest of the diagram shows that there is no perturbation of the UV parameters that could ruin the fractonic IR behaviour (\emph{robustness}).}
    \label{fig:beta_streams}
\end{figure*}

\item It is worth to frame these results within the context of the \emph{robustness} in QFT exposed in \cite{Seiberg:2020}. There, it is commented that usually, in condensed matter systems, the UV models posses not many global internal symmetries $\mathcal{G}_{UV}$, being them typically, $\mathbb{Z}_2$,  $U(1)$, or the trivial one. It is assumed  that by a fine tuning of the short distance parameters, one may find a low-energy theory with an enhanced emergent global symmetry $\mathcal{G}_{IR}$, for instance, a dipole symmetry.  The interesting question is whether the low-energy model will preserve the emergent symmetry $\mathcal{G}_{IR}$ once the short distance parameters are slightly deformed. Our results in Eqs. \eqref{eq:beta_func_ng} suggest that a small deformation of the short-distance system can not wreck the $\mathcal{G}_{IR}$ dipole symmetry at long distances and therefore, we can say that this symmetry is robust. 
\end{enumerate}

\section{Conclusions}
\label{sec:discussion}
 In this paper we have studied the ground state properties and the RG flows of a continuous model of interacting fractons using a nonperturbative variational approach based on the functional Schr\"odinger picture in QFT.

We have found that a Gaussian approximation, while allowing to carry out a consistent RG flow analysis of the theory at 1-loop, only grasps the interaction terms associated to the longitudinal motion of dipoles. 

In order to investigate the true non-Gaussian features of these models, we have proposed a non-Gaussian ansatz wavefunctional based on NLCT.  Without losing sight on the fact our results have been obtained for a concrete NLCT, we argue that those are illustrative and general enough (in the NLCT sense) as to provide a useful kind of understanding on the essence of the nonperturbative physics associated to the models under consideration. Concretely, by providing a variational improved effective potential, the post-Gaussian ansatz is able to capture the transverse part of the the interaction between dipoles in such a way that a non trivial RG flow for this term is obtained and analyzed. This is something that cannot be derived from a pure Gaussian ansatz or large $N$ techniques. Our non-perturbative calculation of the RG flow  suggests that dipole symmetries that manifest due to the strong coupling of dipoles, robustly arise at low energies from UV models without that symmetry. In other words, these symmetries account for an emergent phenomena rather than being well defined properties of microscopic models. 

Some interesting future directions would be to perform explicit optimizations of the improved effective potential in order to study  the phase transition between the broken and unbro-
ken phases of the theory (see \cite{Afxonidis:2023} for a recent work on this line). Indeed, as the the theory is non-Gaussian and strongly coupled close to the critical point transition, even a mean field approximation of the transition is not possible. It  would be interesting to clarify if the phase transition is a second order continuous one or first order. Other interesting future research lines consist in extending these kind of nonperturbative analysis to dipole-symmetric fermionic models, possibly obtained by generalizing \cite{Jensen:2022},
and to apply similar techniques to study interacting models possesing subsystem symmetries.

\section*{Acknowledgements}
I thank Jos\'e Juan Fern\'andez-Melgarejo, Luca Romano and Piotr Sur\'owka for useful and interesting discussions on some topics covered in this work. I would like to thank the financial support of the Spanish Ministerio de Ciencia e Innovaci\'on  PID2021-125700NA-C22 and Programa Recualificaci\'on del Sistema Universitario Espa\~{n}ol 2021-2023.

\newpage
\appendix
\section{Appendix A. Gaussian ansatz for a model with subsystem symmetry}
\label{sec:appA}
To illustrate the UV/IR mixing properties of models with subsystem symmetries we consider  a non-relativistic theory of a real scalar $\phi$. The Lagrangian of the theory is 
\beq
\mathcal{L}
= \frac{1}{2} \left[ (\partial_t \phi )^2 -\mu^2 ( \partial_{\square} \phi )^2 \right] ,
\label{eq:lagrangian_ss}
\enq 
where $\mu$ is a constant, and 
$\partial_{\square}$ is a spatial differential operator defined by
\beq
\partial_{\square}:= \partial_1 \cdots \partial_D .
\enq

The  Hamiltonian of the model in momentum space is
\beq\label{eq:hamilt_ss}
H = \int \Dk\, \left[\frac{1}{2}\, \pi(\vk) \pi(-\vk) + \frac{1}{2}\, \mathbf{K}_{\square}^2\, \phi(\vk) \phi(-\vk)\right]\, ,
\enq
with,
\beq
\mathbf{K}_{\square}^2 = \mu^2\, \left(k_1 \cdots k_D \right)^2 \, .
\enq
Applying the Gaussian ansatz in Eq. \eqref{eq:gauss_ansatz} with $\phi$ real, we obtain
\beq\label{eq:energy_ss}
\mathcal{E}= \int \Dk\,  \frac{1}{8} G^{-1}(\vk) + \frac{1}{2}\, \Sigma(\vk)\, G(\vk) \, 
\enq
with 
\beq
\Sigma(\vk)= \mu^2\, \mathbf{K}_{\square}^2\, .
\enq
The variational problem is solved by $\partial \mathcal{E}/\partial G(\vk) =0$ that yields
\beq
G(\vk)=\frac{1}{2\, \sqrt{\Sigma(\vk)}}\, .
\enq
Thus the energy of the ground state energy reads
\beq
\mathcal{E}=\frac{1}{2}\int\, \Dk \,\omega(\vk)\, ,
\enq
with $\omega(\vk)^2 = \Sigma(k)=\mu^2\, \mathbf{K}_{\square}^2$. We see the strong UV/IR mixing of the model as low-energy mixing among small and high momenta components $k_i$ in $\mathbf{K}_{\square}$.

\section{Appendix B. Self-Energy $\Sigma(\vk)$ explicit solutions}
\label{sec:appB}
In this appendix, following \cite{Jensen:2022}, a method to find numerical solutions to $\Sigma(\vk)$ is presented. To this end, we first introduce the generalization of the Gaussian variational method to finite temperature \cite{Lee94}. This will also provide a regularization procedure that enables to find self consistent solutions of $\Sigma(\vk)$.

\paragraph*{Finite Temperature}
We have shown that for the Gaussian ansatz,  the energy density reads as 
\beq\label{eq:energy_densityG}
\mathcal{E}= \frac{1}{8}\int\, \Dk\, G^{-1}(\vk)\, .
\enq
The procedure to obtain the free energy density at a finite inverse temperature $\beta$ is simple and can be extensively discussed in \cite{Lee94, ritschel90}. For the Gaussian approach to the model in \eqref{eq:fracton_lagrangian} this amounts to,
\beq\label{eq:free_energy}
\mathcal{F}_0 = \frac{1}{8}\int\, \Dk\, \Bigg[G^{-1}(\vk)\, + \frac{2}{\beta}\log \left[1 -\exp\left(-\beta G^{-1}(\vk)\right)\right]\Bigg]\, ,
\enq
which tends to the ground state values as $\beta \to \infty$ as well as $\mathcal{F}_0$ asymptotes to \eqref{eq:energy_densityG} . 

\begin{figure*}
    \centering
    \includegraphics[width = 0.97\linewidth]{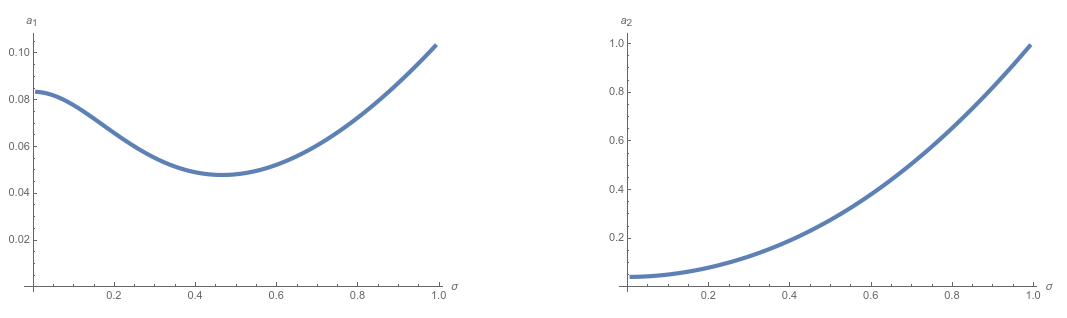}
    \caption{Self-energy $\Sigma_{\sigma}(\vk)$ parameters as a function of $\sigma$ for $D=3$,  $m = 0.1$, $\gamma=0.3$ and $\lambda= 1$. The low temperature limit can be achieved numerically for $\beta \sim 3$.}
    \label{fig:sigma_param2}
\end{figure*}

Operationally speaking, the expectation value of any quantity at finite temperature can be computed using the replacements
\beq\begin{split}\label{eq:finite_T_kernels}
G(\vk) & \to G(\vk)\, \coth\left(\frac{\beta\, G^{-1}(\vk)}{2} \right) = G(\vk)\, \left(1 +\frac{2}{e^{\beta\, G^{-1}(\vk)}-1} \right) \, \\
\frac{1}{G(\vk)} & \to \frac{1}{G(\vk)} \, \coth\left(\frac{\beta\, G^{-1}(\vk)}{2} \right) = G^{-1}(\vk)\, \left(1 +\frac{2}{e^{\beta\, G^{-1}(\vk)}-1} \right)\, ,
\end{split}
\enq
where the equality allows a simple separation into IR and UV contributions: the first term is the zero-temperature result, and the second term, which involves the relativistic Bose statistical factor, decreases exponentially at large momentum. Noteworthily, the self- energy at finite temperature reads 
\beq\label{eq:self_energy_beta}
\Sigma(\vk) = 2\, \int \Dk'\, \bar{V}^{(4)}(\vk,\vk')\, G(\vk')\, \coth\left(\frac{\beta\, G^{-1}(\vk')}{2} \right)\, .
\enq
\paragraph*{$\Sigma(k)$ explicit solutions}
From \eqref{eq:self_energy_beta}, we write a regularized version of the the self-energy \eqref{eq:self_energy_beta} as
\beq\begin{split} \label{eq:Dyson2_reg}
\Sigma(\vk)&=\frac{s_D}{2}\, \int dq\, q^{D-1}\, \bar{V}^{(4)}(\vk,\vq)\, \frac{\left[\coth\left(\beta\left(2\Sigma(\vq) + m^2 \right)^{1/2}\right)-1\right]}{\left(2\Sigma(\vq) + m^2 \right)^{1/2}}\, ,\\
s_D &= \frac{1}{\Gamma(D/2) (2 \pi)^{D/2}}\, .
\end{split}
\enq
where the vertex $\bar{V}^{(4)}(\vk,\vq)=\frac{\lambda_S}{4}\, \vert |\vk -\vq|^2 + 2\gamma \vert^2$ is a polynomial in $|\vk|$ of degree 4 and so $\Sigma(\vk)$ is too. The regularization above effectively removes contributions of high energy modes as $\coth(\beta x) \sim 1$ as $\beta x \to \infty$, that is, the prescription removes the contributions of high energy momentum modes.
As the theory in Eq \eqref{eq:fracton_lagrangian} is rotationally invariant, we make a rotationally invariant ansatz 
\beq
\Sigma(\vk) = a_0 + a_1 |\vk|^2+a_2 |\vk|^4\, .
\enq
Plugging this ansatz into \eqref{eq:Dyson2_reg} and picking the terms in the vertex that yield rotationally invariant terms, one obtains  
\beq\begin{split}\label{eq:param} 
a_0 & = a_2\, \frac{\gamma^2}{2} + \frac{D\gamma}{2(D+2)} (a_1 -2\gamma a_2)+ \frac{s_D\, \lambda}{2}\, \int dq\, q^{D+3}\, \frac{\left[\coth\left(\beta\left(2a_2 |\vq|^{4} + 2 a_1 |\vq|^{2} +2 a_0+ m^2 \right)^{1/2}\right)-1\right]}{\left(2a_2 |\vq|^{4} + 2 a_1 |\vq|^{2} +2 a_0+ m^2 \right)^{1/2}}\, ,\\
a_1 & = \gamma\, a_2\frac{(D+2)}{D} + s_D\, \lambda\, \int dq\, q^{D+1}\, \frac{\left[\coth\left(\beta\left(2a_2 |\vq|^{4} + 2 a_1 |\vq|^{2} +2 a_0+ m^2 \right)^{1/2}\right)-1\right]}{\left(2a_2 |\vq|^{4} + 2 a_1 |\vq|^{2} +2 a_0+ m^2 \right)^{1/2}}\, ,\\
a_2 & = \sigma^{2}\lambda\, + \frac{s_D\, \lambda}{2}\, \int dq\, q^{D-1}\, \frac{\left[\coth\left(\beta\left(2a_2 |\vq|^{4} + 2 a_1 |\vq|^{2} +2 a_0+ m^2 \right)^{1/2}\right)-1\right]}{\left(2a_2 |\vq|^{4} + 2 a_1 |\vq|^{2} +2 a_0+ m^2 \right)^{1/2}}\, ,
\end{split}
\enq
with the constraint $\sigma\, a_0 =0$ and $\lambda = \lambda_S/4$. This amounts to a set of coupled equations for the parameters $(a_0,a_1,a_2, \sigma)$ that can be solved  numerically. In this work we have been mainly interested in solving the equations for $\sigma \neq 0$ and $\beta \to \infty$. We show some of these solutions in Fig \ref{fig:sigma_param2}. 

\section{Appendix C. Optimization of the post-Gaussian ansatz}
\label{sec:appC}
\subsection*{Optimization of $c(\vp,\vq,\vr)$}
The post-Gaussian energy expectation value is given in \eqref{eq:ng_energy_vev} and here we conveniently rewritte it as
\beq\begin{split}\label{eq:ng_energy_vev_alt}
\mathcal{E}_{NG}&=\mathcal{E}_G -2\alpha\, \tilde{\chi}^{(1)} + 4\, \alpha^2\, \tilde{\chi}^{(2)}\, ,\\
\tilde{\chi}^{(1)} &= \int_{\vp,\vq,\vr}\, \left[\Gamma^{(4)}_1(\vp,\vq,\vr) + \Gamma^{(4)}_1(-\vp,\vq,\vr)\right]\, G(\vp)\, G(\vq)\, G(\vr)\, ,\\
\tilde{\chi}^{(2)} & = \int_{\substack{\vp_1,\vp_2\\ \vq_1,\vq_2}}\, \Gamma^{(4)}_2(\vp_1,\vp_2,\vq_1,\vq_2) \,  G(\vp_1) G(\vp_2) G(\vq_1) G(\vq_2) \,  .
\end{split}
\enq

As commented in the main text, we choose $\alpha$ in such a way that 
\beq
\alpha\, \lambda_S\ll 1\, \quad \alpha\, \lambda_T\ll 1\, ,
\enq
with the coupling constants $\lambda_T$  and $\lambda_S$ not necessarily small. With this choice, the optimization equations greatly simplify as $G(\vk)$ and $\Sigma(\vk)$ can be approximated at $\mathcal{O}(\alpha)$ to the Gaussian solution in Eq. \eqref{eq:Sigma_condensate}. This leaves the optimization of the variational parameters $c(\vp,\vq,\vr)$ as a decoupled process that can be carried out by doing
\beq
\frac{\delta\, \mathcal{E}_{NG}}{\delta\, c(\vp,\vq,\vr)} =  -2\alpha\, \frac{\delta\, \tilde{\chi}^{(1)}}{\delta\, c(\vp,\vq,\vr)}  + 4\, \alpha^2\, \frac{\delta\, \tilde{\chi}^{(2)}}{\delta\, c(\vp,\vq,\vr)} = 0\, .
\enq
As a result one obtains the integral equation
\beq\label{eq:int_eq_optim}
2\int_{\vr}\, V^{(4)}(-\vr,-\vp,\vq,\vp-\vq+\vr)\left[\tilde{c}(-\vq,-\vp,\vr)+ \tilde{c}(-\vq,\vp,\vr)\right]\, G(\vr)  = \bar{V}^{(4)}(\vp,\vq)\, ,
\enq
which can also be  written as
\beq\label{eq:int_eq_optim2}
4\int_{\vq,\vr}\, V^{(4)}(-\vr,-\vp,\vq,\vp-\vq+\vr)\left[\tilde{c}(-\vq,-\vp,\vr)+ \tilde{c}(-\vq,\vp,\vr)\right]\, G(\vq)\, G(\vr)  = \Sigma(\vp)\, ,
\enq

where, noting that $\alpha$ amounts to a normalization value for the parameters $c(\vp,\vq,\vr)$ that we use as a tracking parameter, solutions to the equation above refer to the fully meaningful quantity $\tilde{c}(\vp,\vq,\vr) = \alpha\, c(\vp,\vq,\vr)$. Analytical solutions to this equation may be challenging to obtain, even approximately. In this situation, it thus difficult to gain any insight on the structure of these parameters. Fortunately, it is possible to make advance by using a simplified functional form for the variational parameters $c(\vp,\vq,\vr)$.

\subsection*{Functional form of $f(\vp,\vq_1,\vq_2,\vq_3)$}
A suitable way of accomplishing \eqref{eq:constraint} is decomposing  
\beq
f(\vp,\vq_1,\vq_2,\vq_3)\ = g(|\vp|,|\vq_1|,|\vq_2|,|\vq_3|)\, \mathbf{L}(\vp)\, \mathbf{H}(\vq_1)\,   \mathbf{H}(\vq_2)\, \mathbf{H}(\vq_3)\, ,
\label{eq:f-variational}
\enq
where $g(|\vp|,|\vq_1|,|\vq_2|,|\vq_3|)$ is scalar function to be determined by energy minimization and we have imposed that $\mathbf{L}(\vp)\cdot \mathbf{H}(\vp) = 0$, that is, the domains of momenta where $\mathbf{L}$ and $\mathbf{H}$ are different from zero must be disjoint, up to sets of measure zero. Refs \cite{polley89, ritschel90} provide a useful functional form  for $\mathbf{L}$ and $\mathbf{H}$  as
\beq
\begin{split}\label{eq:cutoff_ansatz}
\mathbf{L}(\vk)&=\Gamma \left[\left(\frac{|\vk|}{\Delta_0}\right)^2\right]\, , \\ 
\mathbf{H}(\vk)&= \left(\Gamma \left[\left(\frac{\Delta_0}{|\vk|}\right)^2\right]-\Gamma \left[\left(\frac{\Delta_1}{|\vk|}\right)^2\right]\right)\, ,
\end{split}
\enq
with  $\Delta_{0}<\Delta_1$ being two variational and coupling dependent momentum cutoffs and  $\Gamma(x)\equiv \theta(1-|x|)$ with $\theta(x)$ the Heaviside step function. On very general grounds, $f(\vp,\vq_1,\vq_2,\vq_3)$ can be understood as separating the Fourier components of the field $\phi$ into non-overlapping domains of ``high''-$\mathbf{H}$ and "low"-$\mathbf{L}$- momenta. The variational parameters $\Delta$ determine the size of these non-overlapping regions in momentum space. The effect of the functional form of $f(\vp,\vq_1,\vq_2,\vq_3)$ on, for instance, Eq. \eqref{eq:nlct1} is to shift the "low"-$\mathbf{L}$- momentum modes of $\phi$ by a nonlinear function of the ``high''-$\mathbf{H}$ modes, providing therefore, once optimized, a variational integration of the effects of the high energy modes into the low energy physics of the problem.

With this, and noting that the main contribution for a post-Gaussian improvement of the ground state energy is given by $\tilde{\chi}^{(1)}$ in \eqref{eq:ng_energy_vev_alt}, a (sub-)optimal way of finding the variational parameters $f(\vp,\vq_1,\vq_2,\vq_3)$ consist in \cite{polley89, ritschel90, Qian:2022} fixing $g(|\vp|,|\vq_1|,|\vq_2|,|\vq_3|) = 1$ and then numerically obtaining the pair of values for $(\Delta_0,\, \Delta_1)$ that maximize $\tilde{\chi}^{(1)}$.

\section{Appendix D. Other possible NLCT}
\label{sec:appD}
Other NLCT transformations $B$ can be chosen to build the improved post-Gaussian energy expectation value $\mathcal{E}_{NG}$ and the effective potential $\mathbb{V}_{\rm eff}[\sigma]$ to study the model in \eqref{eq:fracton_lagrangian}. In this Appendix we show an example given by the operator
\beq\begin{split}\label{eq:generator_alt}
B&= -\alpha\int_{\vp,\vq,\vr} \, f(\vp,\vq,\vr)\, \frac{\delta}{\delta \phi(-\vp)}\, \bar{\phi}(\vq)\, \phi(\vr)\, \D^{D}(\vp + \vq +\vr)\, \\
&= -\alpha\int_{\vp,\vq,\vr} \, f(\vp,\vq,\vr)\, \pi(\vp)\, \bar{\phi}(\vq)\, \phi(\vr)\, \D^{D}(\vp + \vq +\vr)\, ,
\end{split}
\enq
where $f(\vp,\vq,\vr)$ fulfill Eq. \eqref{eq:constraint}. These constraints imply that the action on the canonical field operators $\phi(\vk),\, \bar{\phi}(\vk)$ and $\pi(\vk),\, \bar{\pi}(\vk)$ is given by
\beq\begin{split}\label{eq:phi_shift}
\phi(\vk) & \to \phi(\vk) + \alpha\, \Phi(\vk)\, \\
\bar{\phi}(\vk) & \to \bar{\phi}(\vk)+ \alpha\, \bar{\Phi}(\vk)\,\, 
\end{split}
\enq
with
\beq\begin{split}\label{eq:phi_shift_explicit}
\Phi(\vk) &= -\int_{\vp,\vq}\, f(\vk,\vp,\vq)\, \bar{\phi}(\vp)\, \phi(\vq)\, \D^{D}(\vk-\vp-\vq)\, ,\\
\bar{\Phi}(\vk) & = 0\, ,
\end{split}
\enq

and
\beq\begin{split}\label{eq:pi_shift}
\pi(\vk) & \to \pi(\vk)+ \alpha\, \Pi(\vk)\, , \\
\bar{\pi}(\vk) & \to \bar{\pi}(\vk) + \alpha\, \bar{\Pi}(\vk)\, 
\end{split}
\enq
with
\beq\begin{split}\label{eq:pi_shift_explicit}
\Pi(\vk) &=  \int_{\vp,\vq}\, f(\vp,\vq,\vk)\, \pi(\vp)\, \bar{\phi}(\vq)\,  \D^{D}(\vk-\vp-\vq)\\
\bar{\Pi}(\vk) &=  \int_{\vp,\vq}\, f(\vp,\vk,\vq)\, \pi(\vp)\, \phi(\vq)\, \D^{D}(\vk-\vp-\vq)\, .
\end{split}
\enq 

Being $\mathcal{U}=e^B$ unitary, one can check that Eqs. \eqref{eq:pi_shift}-\eqref{eq:pi_shift_explicit} ensure that the canonical commutation relations (CCR) still hold under the nonlinear transformed fields.

\newpage

\bibliographystyle{utphys}

\begin{thebibliography}{99}
\bibitem{Chamon:2004}
C.~Chamon,
``Quantum Glassiness,''
Phys. Rev. Lett. \textbf{94}, no.4, 040402 (2005)
doi:10.1103/physrevlett.94.040402
[arXiv:cond-mat/0404182 [cond-mat.str-el]].

\bibitem{Haah:2011}
J.~Haah,
``Local stabilizer codes in three dimensions without string logical operators,''
Phys. Rev. A \textbf{83}, no.4, 042330 (2011)
doi:10.1103/physreva.83.042330
[arXiv:1101.1962 [quant-ph]].

\bibitem{Vijay:2015}
S.~Vijay, J.~Haah and L.~Fu,
``A New Kind of Topological Quantum Order: A Dimensional Hierarchy of Quasiparticles Built from Stationary Excitations,''
Phys. Rev. B \textbf{92}, no.23, 235136 (2015)
doi:10.1103/PhysRevB.92.235136
[arXiv:1505.02576 [cond-mat.str-el]].

\bibitem{Nand:2018}
R.~M.~Nandkishore and M.~Hermele,
``Fractons,''
Ann. Rev. Condensed Matter Phys. \textbf{10}, 295-313 (2019)
doi:10.1146/annurev-conmatphys-031218-013604
[arXiv:1803.11196 [cond-mat.str-el]].

\bibitem{Pretko:2020}
M.~Pretko, X.~Chen and Y.~You,
``Fracton Phases of Matter,''
Int. J. Mod. Phys. A \textbf{35}, no.06, 2030003 (2020)
doi:10.1142/S0217751X20300033
[arXiv:2001.01722 [cond-mat.str-el]].

\bibitem{Slagle:2017}
K.~Slagle and Y.~B.~Kim,
``Quantum Field Theory of X-Cube Fracton Topological Order and Robust Degeneracy from Geometry,''
Phys. Rev. B \textbf{96} (2017) no.19, 195139
doi:10.1103/PhysRevB.96.195139
[arXiv:1708.04619 [cond-mat.str-el]].

\bibitem{Seiberg:2020}
N.~Seiberg and S.~H.~Shao,
``Exotic Symmetries, Duality, and Fractons in 2+1-Dimensional Quantum Field Theory,''
SciPost Phys. \textbf{10} (2021) no.2, 027
doi:10.21468/SciPostPhys.10.2.027
[arXiv:2003.10466 [cond-mat.str-el]].

\bibitem{Seiberg:2020b}
N.~Seiberg and S.~H.~Shao,
``Exotic $U(1)$ Symmetries, Duality, and Fractons in 3+1-Dimensional Quantum Field Theory,''
SciPost Phys. \textbf{9}, no.4, 046 (2020)
doi:10.21468/SciPostPhys.9.4.046
[arXiv:2004.00015 [cond-mat.str-el]].

\bibitem{McGreevy:2022}
J.~McGreevy,
``Generalized Symmetries in Condensed Matter,''
doi:10.1146/annurev-conmatphys-040721-021029
[arXiv:2204.03045 [cond-mat.str-el]].

\bibitem{Pretko:2018}
M.~Pretko,
``The Fracton Gauge Principle,''
Phys. Rev. B \textbf{98}, no.11, 115134 (2018)
doi:10.1103/PhysRevB.98.115134
[arXiv:1807.11479 [cond-mat.str-el]].

\bibitem{Jensen:2022}
K.~Jensen and A.~Raz,
``Large $N$ fractons,''
[arXiv:2205.01132 [hep-th]].

\bibitem{Islam:2023}
M.M.~Islam, K.~Sengupta and R.~Sensarma
"Non-equilibrium dynamics of bosons with dipole symmetry: A large $N$ Keldysh approach"
[arXiv:2305.13372 [cond-mat.quant-gas]].      ,
     
\bibitem{Li:2021}
H.~Li and P.~Ye,
``Renormalization group analysis on emergence of higher rank symmetry and higher moment conservation,''
Phys. Rev. Res. \textbf{3}, no.4, 043176 (2021)
doi:10.1103/PhysRevResearch.3.043176
[arXiv:2104.03237 [cond-mat.quant-gas]].

\bibitem{Distler:2021}
J.~Distler, M.~Jafry, A.~Karch,and A.~Raz,
``Interacting fractons in 2+1-dimensional quantum field theory,''
JHEP \textbf{03}, 070 (2022)
[erratum: JHEP \textbf{03}, 115 (2023)]
doi:10.1007/JHEP03(2022)070

\bibitem{Han:2022}
S.~Han and Y.B.~Kim	
"Non-Fermi liquid induced by Bose metal with protected subsystem symmetries"
Phys.Rev.B \textbf{106} (2022) no.8 l081106
doi:10.1103/physrevb.106.l081106
[arXiv:2102.05052 [cond-mat.stat-mech]].


\bibitem{Grosvenor:2022}
K.~T.~Grosvenor, R.~Lier and P.~Sur\'owka,
``Fractonic Berezinskii-Kosterlitz-Thouless transition from a renormalization group perspective,''
Phys. Rev. B \textbf{107} (2023) no.4, 045139
doi:10.1103/PhysRevB.107.045139
[arXiv:2207.14343 [cond-mat.stat-mech]].

\bibitem{Pretko:2017}
M.~Pretko, L.~ Radzihovsky
"Fracton-Elasticity duality"
Phys. Rev. Lett. \textbf{120}, 195301 (2018)
doi:10.1103/PhysRevLett.120.195301
[arXiv:1711.11044v2 [cond-mat.str-el]].

\bibitem{Doshi:2020}
D.~Doshi and A.~Gromov,
``Vortices and Fractons,''
[arXiv:2005.03015 [cond-mat.str-el]].

\bibitem{Zechmann:2022}
P.~Zechmann, E.~Altman, M.~Knap and J.~Feldmeier,
``Fractonic Luttinger liquids and supersolids in a constrained Bose-Hubbard model,''
Phys. Rev. B \textbf{107}, no.19, 195131 (2023)
doi:10.1103/PhysRevB.107.195131
[arXiv:2210.11072 [cond-mat.quant-gas]].

\bibitem{Hatfield:1992}
B.~Hatfield,
``Quantum field theory of point particles and strings,''
Advanced Book Program, Westview Press, 1992,
ISBN 0-201-11982-X

\bibitem{Symanzik:1981}
K.~Symanzik,
``Schrodinger Representation and Casimir Effect in Renormalizable Quantum Field Theory,''
Nucl. Phys. B \textbf{190}, 1-44 (1981)
doi:10.1016/0550-3213(81)90482-X


\bibitem{polley89} L. Polley and U. Ritschel, 
"Second Order Phase Transition in $\lambda\phi^4$ in Two-dimensions With Non-gaussian Variational Approximation", 
\textit{Phys. Lett.} B 221,44 (1989).

\bibitem{ritschel90} U. Ritschel, 
"Improved effective potential by nonlinear canonical transformations", 
\textit{Zeitschrift für Physik} C 47(3):457-467 (1990).

\bibitem{ibanez} R. Iba\~nez-Meier, A. Mattingly, U. Ritschel, and P. M. Stevenson, 
"Variational calculations of the effective potential with non-Gaussian trial wave functionals", 
\textit{Phys.Rev.}  D45, (1992) 15.

\bibitem{Jain:2023}
A.~Jain, K.~Jensen, R.~Liu and E.~Mefford,
``Dipole superfluid hydrodynamics,''
[arXiv:2304.09852 [hep-th]].

\bibitem{Bidussi:2021}
L.~Bidussi, J.~Hartong, E.~Have, J.~Musaeus and S.~Prohazka,
``Fractons, dipole symmetries and curved spacetime,''
SciPost Phys. \textbf{12}, no.6, 205 (2022)
doi:10.21468/SciPostPhys.12.6.205
[arXiv:2111.03668 [hep-th]].

\bibitem{Coleman:1973}
S.~R.~Coleman and E.~J.~Weinberg,
``Radiative Corrections as the Origin of Spontaneous Symmetry Breaking,''
Phys. Rev. D \textbf{7}, 1888-1910 (1973)
doi:10.1103/PhysRevD.7.1888

\bibitem{Barnes:1980}
Barnes, Ted, and G. I. Ghandour. ``Renormalization of trial wave functionals using the effective potential." \textit{Physical Review D} 22.4 (1980): 924.

\bibitem{Seiberg:2021}
P.~Gorantla, H.~T.~Lam, N.~Seiberg and S.~H.~Shao,
``Low-energy limit of some exotic lattice theories and UV/IR mixing,''
Phys. Rev. B \textbf{104}, no.23, 235116 (2021)
doi:10.1103/PhysRevB.104.235116
[arXiv:2108.00020 [cond-mat.str-el]

\bibitem{Zhou:2021}
Z.~Zhou, X.~F.~Zhang, F.~Pollmann and Y.~You,
``Fractal Quantum Phase Transitions: Critical Phenomena Beyond Renormalization,''
[arXiv:2105.05851 [cond-mat.str-el]].

\bibitem{Lake:2021}
E.~Lake,
``Renormalization group and stability in the exciton Bose liquid,''
Phys. Rev. B \textbf{105}, no.7, 075115 (2022)
doi:10.1103/PhysRevB.105.075115
[arXiv:2110.02986 [cond-mat.str-el]].

\bibitem{Stevenson:1985}
P.M. Stevenson,
``The Gaussian Effective Potential. 2. Lambda phi**4 Field Theory,''
\textit{Phys.\ Rev.\ D} {\bf 32} (1985) 1389.
doi:10.1103/PhysRevD.32.1389

\bibitem{Kogan:1994}
I.~I.~Kogan and A.~Kovner,
``Variational approach to the QCD wave functional: Dynamical mass generation and confinement,''
Phys. Rev. D \textbf{52}, 3719-3734 (1995)
doi:10.1103/PhysRevD.52.3719
[arXiv:hep-th/9408081 [hep-th]].

\bibitem{Melgarejo:2020}
J.~J.~Fernandez-Melgarejo and J.~Molina-Vilaplana,
``Non-Gaussian Entanglement Renormalization for Quantum Fields,''
JHEP \textbf{07}, 149 (2020)
doi:10.1007/JHEP07(2020)149
[arXiv:2003.08438 [hep-th]].

\bibitem{Melgarejo:2020b}
J.~J.~Fernandez-Melgarejo and J.~Molina-Vilaplana,
``Entanglement Entropy: Non-Gaussian States and Strong Coupling,''
JHEP \textbf{02}, 106 (2021)
doi:10.1007/JHEP02(2021)106
[arXiv:2010.05574 [hep-th]].

\bibitem{Melgarejo:2019}
J.~J.~Fernandez-Melgarejo, J.~Molina-Vilaplana and E.~Torrente-Lujan,
``Entanglement Renormalization for Interacting Field Theories,''
Phys. Rev. D \textbf{100}, no.6, 065025 (2019)
doi:10.1103/PhysRevD.100.065025
[arXiv:1904.07241 [hep-th]].

\bibitem{Melgarejo:2021}
J.~J.~Fernandez-Melgarejo and J.~Molina-Vilaplana,
``The large N limit of icMERA and holography,''
JHEP \textbf{04}, 020 (2022)
doi:10.1007/JHEP04(2022)020
[arXiv:2107.13248 [hep-th]].

\bibitem{Qian:2022}
T.~Qian, J.~J.~Fernandez-Melgarejo, D.~Zueco and J.~Molina-Vilaplana,
``Non-Gaussian variational wave functions for interacting bosons on a lattice,''
Phys. Rev. B \textbf{107}, no.3, 035121 (2023)
doi:10.1103/PhysRevB.107.035121
[arXiv:2211.04320 [cond-mat.str-el]].

\bibitem{ritschel:91} U. Ritschel, 
"Renormalization of the post-Gaussian effective potential", 
\textit{Zeitschrift für Physik} C 51:469-475 (1991).

\bibitem{Afxonidis:2023}
E.~Afxonidis, A.~Caddeo, C.~Hoyos and D.~Musso,
``Dipole symmetry breaking and fractonic Nambu-Goldstone mode,''
[arXiv:2304.12911 [hep-th]].

\bibitem{Lee94}
H.~J.~Lee, K.~Na and J.~H.~Yee,
``Finite temperature quantum field theory in the functional Schrodinger picture,''
Phys. Rev. D \textbf{51}, 3125-3128 (1995)
doi:10.1103/PhysRevD.51.3125
\end{thebibliography}

\providecommand{\href}[2]{#2}\begingroup\raggedright
\endgroup

\end{document}